\documentclass[reprint,amsmath, twocolumn, amssymb, aps, notitlepage, superscriptaddress, prx]{revtex4-2}
\usepackage{graphicx}
\usepackage{dcolumn}
\usepackage{bm}
\usepackage[mathlines]{lineno}
\usepackage{physics}
\usepackage{float}
\usepackage{xcolor}
\usepackage[colorlinks=true, citecolor=blue, urlcolor=blue]{hyperref}
\usepackage[normalem]{ulem}
\usepackage{natbib}

\newcommand{\ncom}[1]{{{#1}}}
\newcommand{\com}[1]{{{#1}}}

\newcommand{\rahul}[1]{{\color[rgb]{0.6,0.0,0.6}{#1}}}

\begin{document}

\title{Sensing atomic superfluid rotation beyond the standard quantum limit}

\author{Rahul Gupta} \affiliation{Department of Physics, Indian Institute of Technology Bombay, Powai, Mumbai 400076, India}

\author{Pardeep Kumar} 
\affiliation{Max Planck Institute for the Science of Light, Staudtstra{\ss}e 2, 91058 Erlangen, Germany}

\author{Rina Kanamoto} 
\affiliation{Department of Physics, Meiji University, Kawasaki, Kanagawa 214-8571, Japan}

\author{M. Bhattacharya}
\affiliation{School of Physics and Astronomy, Rochester Institute of Technology,
84 Lomb Memorial Drive, Rochester, New York 14623, USA}

\author{Himadri Shekhar Dhar} 
\affiliation{Department of Physics, Indian Institute of Technology Bombay, Powai, Mumbai 400076, India}
\affiliation{Centre of Excellence in Quantum Information, Computation, Science and Technology, Indian Institute of Technology Bombay, Mumbai 400076, India}

\date{\today}    

\begin{abstract}
Atomic superfluids formed using Bose-Einstein condensates (BECs) in a ring trap are currently investigated in the context of superfluid hydrodynamics, quantum sensing and matter-wave interferometry. The characterization of the rotational properties of such superfluids is important, but can presently only be performed by using optical absorption imaging, which completely destroys the condensate. Recent studies have proposed coupling the ring BEC to optical cavity modes carrying orbital angular momentum to make minimally destructive measurements of the condensate rotation. The sensitivity of these proposals, however, is bounded below by the standard quantum limit set by the combination of laser shot noise and radiation pressure noise. In this work, we provide a theoretical framework that exploits the fact that the interaction between the scattered modes of the condensate and the light reduces to effective optomechanical equations of motion. We present a detailed theoretical analysis to demonstrate that the use of squeezed light and backaction evasion techniques allows the angular momentum of the condensate to be sensed with noise well below the standard quantum limit. Our proposal is relevant to atomtronics, quantum sensing and quantum information. 

\end{abstract}

\maketitle

\section{Introduction\label{into}}

Quantum sensing with highly controllable ultracold atoms \cite{Lewenstein2007,Bloch2008}
is an actively developing field that provides new pathways to investigate foundational aspects of modern physics such as precise measurements of theoretically predicted fundamental constants \cite{Barontini2022,PhysRevA.106.032807} and ultra-precise tracking of time~\cite{Campbell2011,Kyle2021}. 
In recent years, there has been growing interest in the use of atomic superfluids such as Bose-Einstein condensates (BEC) confined inside a toroidal or ring trap
for sensing~\cite{Marti2015,Ragole2016,Pelegri2018}.
Such configurations have been built using magnetic traps~\cite{Ryu2007}, time-averaged adiabatic potentials~\cite{Sherlock2011}, painted dipole potentials~\cite{Ryu2013}, and intersecting ``sheet'' and ``tube'' laser beams~\cite{Beattie2013}. Besides sensing, several fundamental \cite{Ferdinand2008,Ritter2009,Baumann2010} and technologically relevant properties have been demonstrated ranging from persistent currents~\cite{Ryu2007, Beattie2013} and quantum phase slips~\cite{Wright2013,Snizhko2016} to atomic analogs of superconducting quantum interference devices~\cite{Ryu2013} and atomtronic circuits~\cite{Ramanathan2011, AmicoRMP2022}.

While investigating and harnessing the rotation of atomic superfluids are of intrinsic interest to researchers, 
capturing information about
these rotational properties can be challenging. For instance, existing experimental approaches for probing the BEC angular momentum in ring traps completely destroy the condensate and as such do not allow for continuous observation
\cite{Ryu2007,Beattie2013,Wright2013,Marti2015,PhysRevA.73.041605,Mukherjee2022}.
Non-destructive study of (simply connected, i.e. harmonically trapped) condensates ~\cite{Anderson2000} captures the density profile and not the phase - or more rigorously, the phase gradient - of the rotating states, which is necessary for estimating the angular momentum or winding number of the BECs. 
In fact, these methods rely on vortex precession for such measurements, which is not accessible
while studying superfluids trapped in toroidal or ring traps, where the vortex is pinned, i.e. fixed in position, by the 
potential. 

More sophisticated techniques involving ring traps involve detecting the Doppler shift in phonon modes generated in the condensate~\cite{Kumar2016} or using an integrated atomic circuit, where information about the ring condensate is captured by studying the dynamics inside a tunnel-coupled rectilinear guide~\cite{Safaei2019}.
Recent studies~\cite{Kumar2021,Pradhan2023a,Pradhan2023b} have proposed the use of quantum sensing techniques 
based on cavity optomechanics~\cite{Aspelmeyer2014} to detect the rotational properties of superfluids in ring traps. The approach relies on the creation of sidemode excitations of the condensate due to interaction of the atoms with the orbital angular momentum (OAM) of the light~\cite{Allen2003} injected inside the cavity. 
While the above proposal allows for improvement in rotational sensitivity by a few orders of magnitude over other known techniques, it is still impacted by the noise imparted by the measurement setup, which restricts it to within the bounds of the standard quantum limit (SQL) \cite{Aspelmeyer2014,Kim2016}.

In this work, we 
draw upon more powerful machinery from the toolbox of
quantum optomechanics to study atomic superfluids inside a ring trap and show how to achieve rotational sensing 
of the condensate beyond the SQL. 
We consider a setup similar to Ref.~\cite{Kumar2021} where the cavity is driven with a superposition of light beams with OAM $\pm l\hbar$, as shown in
Fig.~\ref{fig:setup_illustration},  
thus giving rise to effective optomechanical interactions between the two sidemodes of the condensate and the optical mode. Additionally, the effect of atomic collisions in the condensate is also taken into account in the system dynamics.

There are three primary active noise sources in such systems \cite{Aspelmeyer2014} that one actively seeks to mitigate. The first is the shot noise associated with light, the second is the effect of radiation pressure created by the act of measurement, which is also known as measurement backaction noise~\cite{Caves1980b,Borkje2010,Purdy2013} and finally, the thermal and quantum fluctuations of the excited modes of the condensate. Using phase-squeezed light as input~\cite{Caves1981,Fabre1994},  the noise can be significantly lowered, especially at low input powers where the shot noise is dominant, thus achieving significantly higher sensitivity in detecting rotational properties of BECs in a toroidal trap. 
However, using squeezed light alone is not sufficient to overcome the radiation pressure or backaction noise, which prevents the noise spectrum from going below the SQL. 
%
%
To overcome this, sophisticated quantum nondemolition measurement techniques are often required~\cite{Thorne1978,Braginsky1980, Kimble2001,Purdy2013b,Kampel2017,Moller2017,Shomroni2019,Ganapathy2023}, which can mitigate the effect of measurement backaction. 
In this work, we consider a quantum nondemolition method based on backaction evasion (BAE) measurements in a hybrid  system~\cite{Braginsky1992,Clerk2010}, consisting of the two BEC sidemodes and the optical cavity field.


For the condensate in a ring trap, the BAE measurement approach is based on driving the cavity with two off-resonant optical fields~\cite{Woolley2013}. The key idea here is for the cavity mode to couple dynamically to only one pair of the collective quadratures of the two condensate modes, which are then measured. However, the measurement backaction only drives the conjugate set of quadratures, which do not affect the dynamics. Going beyond the conventional analysis~\cite{Woolley2013}, we adopt a Floquet theory based approach to obtain the steady states of an optomechanical system using a bichromatic drive. However, unlike the previous studies~\com{\cite{Malz2016a,Malz2016b,Brunelli2019}}, we consider a hybrid system consisting of two sidemodes
of the condensate and the optical cavity, 
and extend the formalism to work in the bad cavity regime. The time-dependent equations of motion are solved beyond the rotating wave approximation (RWA), where counter-rotating terms are present.
Using such a BAE method, we find that nondemolition measurements of the scattered modes can be performed with the noise spectrum going below the SQL. Importantly, we show that by {combining} the BAE scheme with phase-squeezed input light~\cite{Ghosh2022} in the cavity, 
\textcolor{blue}{the}
angular momentum of the condensate can be measured with high sensitivity and noise below the SQL even at higher input powers.

The paper is arranged as follows. In Sec.~\ref{sec:model}, the physical setup of a BEC confined in a toroidal trap, placed inside a cavity is presented. This is followed by Sec.~\ref{sec:squeezed}, where the theoretical framework for rotational sensing using squeezed OAM-carrying modes is presented. In Sec.~\ref{sec:bae}, the backaction evasion method using a bichromatic driving field is introduced in the context of rotational sensing, with derivation of the equations of motion beyond the rotating wave approximation. A summary and conclusions are presented in Sec.~\ref{sec:conc}.

\section{Model\label{sec:model}}

\begin{figure}[!t]
\centering
\includegraphics[width=0.5\textwidth]{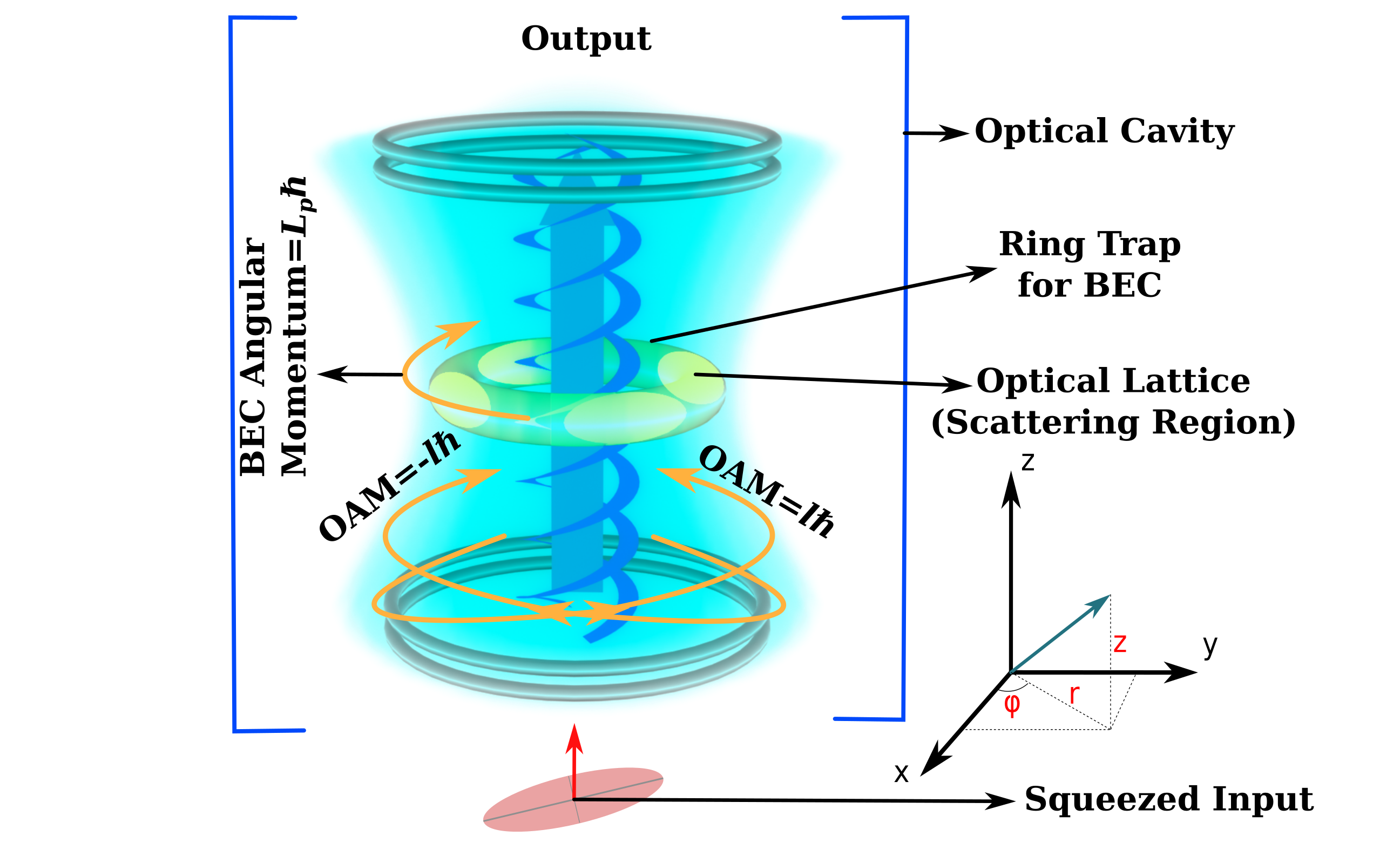}
\caption{
BEC in a toroidal ring trap inside an optical cavity. Superposed laser beams carrying OAM $\pm l\hbar$ (blue corkscrew wavefront and orange lines) are injected inside the cavity (blue lines). This creates an angular optical lattice (green regions) inside the ring trap (cyan toroidal shape) which contains the BEC. The input light is squeezed and shone from the bottom, while the output light after interaction is gathered from the top for homodyne detection and rotational sensing.}
\label{fig:setup_illustration}
\end{figure}

The primary setup consists of a BEC or an atomic superfluid 
of $N$ atoms each of mass $m$ confined in a ring trap~\cite{Ryu2007,Sherlock2011,Ryu2013,Beattie2013}, which is then placed at the center of an optical cavity, as shown in Fig.~\ref{fig:setup_illustration}. The condensate is trapped using an axial harmonic potential $\frac{1}{2}m \omega_z z^2$ in the vertical $z$ direction and an annular ring trap with radius $R$ in the $x$-$y$ plane, with the radial potential $\frac{1}{2}m \omega_r (r-R)^2$. For such a trap,  the dynamics of the condensate can be limited to a single dimension i.e. the unhindered azimuthal motion along the $\varphi$ direction of the ring trap~\cite{Kumar2021,Morizot2006}.

For sensing the BEC rotation, a laser beam of frequency $\omega$ is diffracted and recombined to form a superposed pair of beams with optical angular momenta $\pm l\hbar$~\cite{Naidoo2012}. The cavity with resonant frequency $\omega_a$ and containing the trapped BEC, is then coherently driven by these OAM beams. Importantly, such a driving leads to the formation of an optical lattice around the axis of the cavity in the azimuthal direction (see Fig.~\ref{fig:setup_illustration}).  
If the condensate rotating inside the cavity has a winding number $L_{p}$,
a small number of the atoms will experience first-order Bragg scattering from the optical lattice~\cite{Blakie2001} and attain the final winding numbers $L_{\pm}=L_{p} \pm 2l$, respectively. For a weak lattice \ncom{\cite{Kumar2021}}, the number of atoms undergoing higher-order scattering will be negligible compared to number of atoms contained in the zeroth (unscattered) and first-order scattering sidemodes.
{We note that even the first order scattered sidemodes contain a very small fraction of the total atoms in the condensate. As such, the sidemodes have a negligible chemical potential and the temperature is not significantly affected due to scattering. This is in contrast to cases where splitting of BEC occurs to form new condensed modes~\cite{Langen2013}.}

The condensate is represented by atomic operators $\Phi(\varphi)$, where the angle $\varphi$ is the azimuthal degree of freedom. Following the scattering picture presented above, the atomic operator can be represented in terms of the persistent current ($C$) and the first-order sidemodes ($C_\pm$) such that 
\begin{equation}
\Phi(\varphi) = \frac{e^{iL_{p}\varphi}}{\sqrt{2\pi}} C + \frac{e^{iL_+\varphi}}{\sqrt{2\pi}} C_+ +\frac{e^{iL_-\varphi}}{\sqrt{2\pi}} C_-,
\end{equation}
where the total number of atoms in terms of the bosonic operators is $C^\dag C + C^\dag_+ C_+ + C^\dag_- C_- = N$. Assuming that the sidemodes are sparsely occupied and taking a mean field approximation for the macroscopic $C$ mode, such that $C^\dag C \approx C^*C = N$, one can define the operators $c = C^* C_+/\sqrt{N}$ and $d = C^* C_-/\sqrt{N}$.
The Hamiltonian of the system, using the above operators and written in the frame rotating with the driving frequency, reduces to~\cite{Kumar2021}:
\begin{align}
H/\hbar&= {\Delta}~a^{\dagger}a + \omega_c c^{\dagger} c + \omega_d d^{\dagger} d +  G a^{\dagger}a \left(X_c+X_d\right) \nonumber\\
&-i \eta\left(a-a^{\dagger}\right) +  g \mathcal{Z},
\label{Heff}
\end{align}
where $\omega_{c~(d)} = \left(L_{p}\pm 2l\right)^2 \hbar/2 I$ ($I$ is the moment of inertia of each atom about the center of the ring) and $a^\dag$ and $a$ are the creation and annihilation operators, respectively, of the optical cavity field.
The term ${\Delta}$ denotes the effective detuning between the driving and cavity frequency.
The field injected in the cavity is a single mode superposition of frequency degenerate OAM light, $\ket{+l\hbar} + e^{i\varphi_0}\ket{-l\hbar}$, where the relative phase $\varphi_0$ can be fixed by the experimentalist. The light intensity in the corresponding optical lattice varies as $\cos^2(l\varphi - \varphi_0/2)$, where $\varphi$ is the azimuthal degree of freedom. The scattering with atoms in the BEC only transforms  $\ket{\pm l\hbar}\rightarrow\ket{\mp l\hbar}$, and the optical lattice remains unchanged. This implies no new optical modes are populated. In this work we consider $\varphi_0=0$, which gives us a symmetric superposition of $\ket{\pm l\hbar}$ (see Appendix \ref{sec:A0} for further details).
%
The displacement operators of the atomic side modes are given by
$X_{c}=(c+c^{\dag})/\sqrt{2}$ and $X_{d}=(d+d^{\dag})/\sqrt{2}$, 
where the effective optomechanical interaction between the displacement $X_{c (d)}$ and the optical lattice is given by $G$. 
$\eta$ is the driving amplitude of the input {beam}.
The final term $g \mathcal{Z}$ represents the effect of collisions between atoms in the BEC, where $g$ is the interaction between any two atoms. The term $G$ and operator $\mathcal{Z}$ are derived in Ref.~\cite{Kumar2021}, with $G=g_a^2 \sqrt{N}/2\sqrt{2}\Delta_a$, where $g_a$ is the coupling strength of a photon with a single atom, $\Delta_a$ is the detuning of input beam with the atomic transition level. $\mathcal{Z}$ contains collision operators up to order $N$, which significantly contributes to the dynamics.


\section{Rotational sensing using squeezed light\label{sec:squeezed}}

The few-mode effective Hamiltonian in Eq.~(\ref{Heff}) describes the system consisting of the condensate with a winding number  
$L_p$ and interacting with the optical field carrying OAM $\pm l \hbar$. The dynamics of the system can be written in the form of the quantum Langevin equations \cite{Aspelmeyer2014,Barchielli2015}. They can be obtained by writing the Heisenberg equations of motion for the cavity and the sidemode operators. Below, we express the equations in terms of only the cavity and position quadrature of sidemodes, which resemble a pair of damped oscillators, driven by photons and stochastic noise operators,
\begin{eqnarray}
&\ddot{X}_c+\gamma \dot{X}_c+\Omega_c^2 X_c = -\tilde{\omega}_c G a^{\dagger} a-\mathcal{A} X_d+\omega_c \epsilon_{c,\mathrm{in},} & \label{qle1}\\
[3pt]
&\ddot{X}_d+\gamma \dot{X}_d+\Omega_d^2 X_d = -\tilde{\omega}_d G a^{\dagger} a+\mathcal{A} X_c+\omega_d \epsilon_{d,\mathrm{in}}, & \label{qle2}\\ 
[3pt]
&\dot{a} = \left(i[{\Delta}-G\left(X_c+X_d\right)]-{\kappa}/{2}\right)a + \eta+\sqrt{\kappa}~ a_{\mathrm{in}},& \label{qle3}
\end{eqnarray}
where $\omega_{c~(d)}^2 = (\omega_{c~(d)}+4gN)^2 -4g^2N^2$ and
$\mathcal{A} = 2gN(\omega_c - \omega_d)$ are the modified sidemode frequencies and coupling between them.
The contributions from the interatomic collisions with strength $g$ arise from the operator $\mathcal{Z}$ in Eq.~\eqref{Heff}\cite{Kumar2021}. The damping and cavity loss rates are given by $\gamma$ and $\kappa$, respectively, and the shifted frequencies $\tilde{\omega}_{c(d)} = \omega_{c~(d)} + 2gN$.
The driving term $\eta=\sqrt{\mathcal{P}_{\rm in}\kappa/\hbar\omega_a}$, where $\mathcal{P}_{\rm in}$ is the input laser power,
and $\epsilon_{c(d),\rm in}, a_{\rm in}$ are the input noise operators that take into account the thermal and quantum noise in the atomic sidemodes and the cavity modes, respectively. 

In the Langevin formalism \cite{Aspelmeyer2014,Barchielli2015}, the noise enters the equations in the form of correlation functions, where the mean fluctuations are zero i.e., $\langle \epsilon_{c(d),\rm in}\rangle =\langle a_{\rm in}\rangle = 0$. We introduce squeezing in the optical modes using the input-output formalism, which allows for modeling quantum fluctuations injected into the cavity, in addition to any laser driven coherent state. As such, squeezed vacuum fluctuations $a_{\rm in}$ are added on top of the coherent cavity state $a$, which can be experimentally realised using a combination of beamsplitters and an optical parametric oscillator~\cite{Wu1987,Park2024}.
Now, if the input cavity mode is squeezed, the general correlations of the noise operators are given by~\cite{Aspelmeyer2014,Fabre1994},
\begin{eqnarray}
&\left\langle Q_{\text{in}}(\omega) Q_{\text{in}}\left(\omega^{\prime}\right)\right\rangle =\left[2 N_r+1+M_r+M_r^{*}\right] \pi \delta\left(\omega+\omega^{\prime}\right),&\nonumber \\[3pt]
&\left\langle P_{\text{in}}(\omega) P_{\text{in}}(\omega^\prime)\right\rangle =\left[2 N_r+1-\left(M_r+M_r^{*}\right)\right] \pi \delta\left(\omega+\omega^{\prime}\right),&\nonumber \\ [3pt]
&\left\langle Q_{\text{in}}(\omega) P_{\text{in}}\left(\omega^{\prime}\right)\right\rangle =i\left[1+\left(M_r^{*}-M_r\right)\right] \pi \delta\left(\omega+\omega^{\prime}\right),&\nonumber \\ [3pt]
&\left\langle P_{\text{in}}(\omega) Q_{\text{in}}\left(\omega^{\prime}\right)\right\rangle =i\left[-1+\left(M_r^{*}-M_r\right)\right] \pi \delta\left(\omega+\omega^{\prime}\right),&\nonumber \\[3pt]
&\left\langle\epsilon_{k,\rm{in}}(\omega)\epsilon_{k,\rm{in}}(\omega ')\right\rangle =\mathcal{B}_k\left(\coth\left[\frac{\hbar\omega_{k}}{2k_{B}T_k}\right] + 1\right)
\delta\left(\omega + \omega'\right),&
\label{eq:correlations}
\end{eqnarray}
where $Q_{\text{in}}=(a_{\text{in}}+a_{\text{in}}^\dag)/\sqrt{2}$ and 
$P_{\text{in}}=i(a_{\text{in}}^{\dagger}-a_{\text{in}})/\sqrt{2}$, are the quadrature operators of the incident light.
Here, $\mathcal{B}_k=2\pi\gamma\omega/\omega_k$ and $T_k$ is the temperature of the BEC, where $k = \{c,d\}$.
$N_r$ and $M_r$ are related to squeezing parameters $r$ and $\theta$ and are given by
\begin{eqnarray}
N_r &=& \sinh^2 r + N_{a}\left(\omega_{a}\right)\left(\sinh^2 r + \cosh^2 r\right)\\
M_r &=& e^{ i \theta}\sinh r \cosh r \left[2N_{a}\left(\omega_{a}\right) + 1\right],
\label{eq:squeeze}
\end{eqnarray}
where $N_{a}(\omega_{a})$ is the number of thermal photons, which we have neglected as cavity frequency is in the optical regime.
Moreover, in the absence of squeezing, the terms $N_r$ and $M_r$ are identically zero, 
and the optical noise correlations reduce to $\langle a_{\rm in}(\omega)a_{\rm in}^\dag(\omega')\rangle = 2\pi\delta\left(\omega+\omega^{\prime}\right)$.

For sensing the winding number $L_{p}$ of the BEC inside the ring trap, the central idea is to detect the phase quadrature of the cavity output field, which contains the two frequencies $\Omega_c$ and $\Omega_d$~\cite{Kumar2021}.
A heuristic reading of Eqs.~(\ref{qle1})-(\ref{qle2}) in the absence of any noise and negligible radiation pressure shows that the sideband mode displacements $X_c$ and $X_d$ oscillate at frequencies $\Omega_c$ and $\Omega_d$, which in turn governs the optical field in Eq.~(\ref{qle3}), as the field operator $a$ is  coupled to the two sideband modes. 
A homodyne detection of the output optical phase quadrature $P_{\rm out}(\omega)$ should therefore give a \com{spectral density} $S(\omega)$, which peaks at $\Omega_c$ and $\Omega_d$. 
Now, if $4g^2N^2\ll (\omega_{c~(d)}+4gN)^2$, the spectral peaks are given by $\omega_{c~(d)} \approx \omega_{c~(d)}+ \Delta \omega_{\rm coll}$, where  $\Delta \omega_{\rm coll} = 2gN(2-gN/\omega_{c~(d)})$. As such, by measuring the
gap between the spectral peaks $\Omega_c-\Omega_d = \omega_c-\omega_d = 4L_{p}l\hbar/I$, where $l$ and $I$ are already known, and the  winding number $L_{p}$ of the BEC can be estimated.
The objective now is to find the optimal conditions such that measurement of output spectrum can be performed with minimal noise and high sensitivity. To achieve this, the equations of motion need to be solved to obtain the \com{spectral density} of the output optical field.

 
The steady state solutions of Eqs.~(\ref{qle1})-(\ref{qle3}) are given by $X^s_{c(d)} = -\tilde{\omega}_{c(d)} G|a^s|^2/\Omega^2_{c(d)}-\mathcal{A}X^s_{d(c)}$ and $a^s = -\eta/\alpha$, where $\alpha = i(\Delta - G(X^s_c+X^s_d))-\kappa/2$.
These solutions exhibit bistability, and for studying the spectral distribution and noise sensitivity, it is preferable to work in the monostable regime~\cite{Kumar2021}. A linear response analysis can then be performed around monostable steady states, such that the system operators can be written as $K = K^s + \delta K$, where $K=\{a,X_c,X_d\}$.
Moreover, the analysis can be carried out in the frequency domain by taking Fourier transforms. Linearizing around the steady state, the relations in Eqs.~\eqref{qle1}-\eqref{qle3} can be rewritten in frequency space as $\mathcal{F}\delta u = D u_{\rm in}$, where the matrix $\mathcal{F}$ is given by
\begin{equation}
\mathcal{F}=
\left(
\begin{array}{cccc}
 \chi_a^{-1} & \Delta'  & 0 & 0 \\
 -\Delta'  & \chi_a^{-1} & \sqrt{2} G a_s & \sqrt{2} G a_s \\
 \sqrt{2} G a_s \tilde{\omega }_c & 0 & \chi_c^{-1} & \mathcal{A} \\
 \sqrt{2} G a_s \tilde{\omega }_d & 0 & -\mathcal{A} & \chi_d^{-1} \\
\end{array}
\right),
\label{eq:matrix}
\end{equation}
$\delta u=\left[\delta Q,\delta P,\delta X_{c},\delta X_{d}\right]^{T}$ and $u_{\text{in}}=\left[Q_{\text{in}}, P_{\text{in}},\epsilon_{c,\rm in},\epsilon_{d,\rm in}\right]^{T}$ 
are vectors containing the fluctuation and input noise operators, respectively, such that $\delta Q=(\delta a + \delta a^{\dag})/\sqrt{2}$,$\delta P=i(\delta a^{\dag} - \delta a)/\sqrt{2}$, and  $D={\rm diag}\left[\sqrt{\kappa},\sqrt{\kappa},\omega_c,\omega_d\right]$ is a diagonal matrix with the system loss rates as elements. $\chi_a^{-1} (\omega)={\kappa}/{2} - i \omega$ and 
$\chi_{c(d)}^{-1} (\omega)=\omega_{c~(d)}^2 - i\omega\gamma -\omega^2$, are susceptibility functions.
%
%
The above matrix equation can be inverted to obtain the required analytic and numerical solutions for $\delta P(\omega)$ and $\delta Q(\omega)$. 
Now, using the input-output formalism \cite{Gardiner2004,Clerk2010}, the output optical quadrature can be found from the relation, 
$\delta K_{\rm out}(\omega)=\sqrt{\kappa}\delta K(\omega) - \delta K_{\rm in}(\omega)$, for $K = \{Q,P\}$. \com{Note that the prefix $\delta$ in front of input and output terms is dropped henceforth, for compactness.}
The analytical expressions for these coefficients are provided in Appendix~\ref{sec:A1}.  

\begin{figure}[t]
\centering
\includegraphics[width=3in]{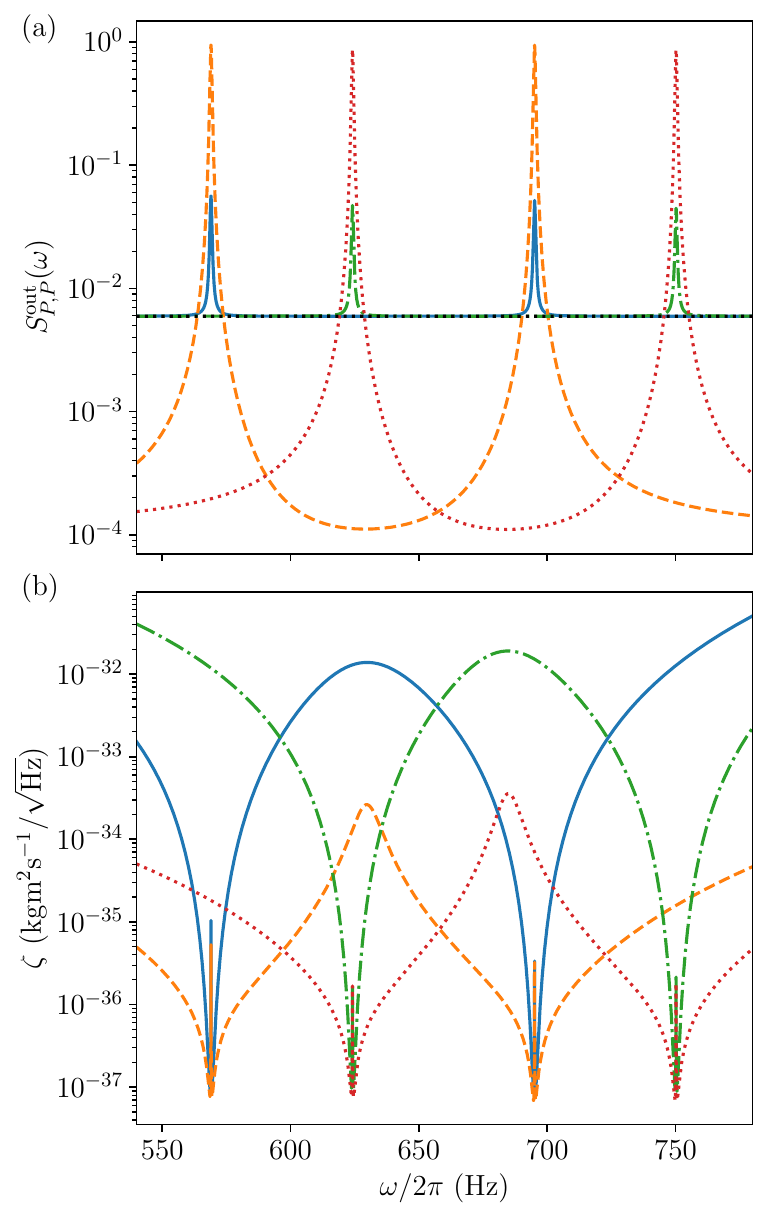}
\caption{Variation of spectral density and sensitivity with frequency. (a) The plot shows the output optical noise spectral density for unsqueezed ($r=0$, solid blue line) and squeezed ($r=2,\theta=\pi$, dashed orange line) input, without inter-atomic collisions ($g = 0$). Cases with inter-atomic interactions ($g \neq 0$) are indicated with green dash-dot line for $r=0$ and with red dotted curve for $r=2,\theta=\pi$. For squeezing angle $\theta = \pi$, the noise spectrum falls below the shot noise level (black-dotted horizontal line). b) Shows the variation of sensitivity for the same set of parameters above. The input power $\mathcal{P}_{\rm in} = 12.4$~fW and the detection angle $\phi=\pi/2$. 
The parameters $N=10^4$, $R=12~\mu$m, $gN \approx 14$~Hz, $I/\hbar = 0.0505$~Hz, and $G/2\pi=7.5$~kHz, correspond primarily to a condensate of sodium atoms~\cite{Kumar2021}.
Other parameters are $\omega_a/2\pi = 10^3$~THz, $\omega_r/2\pi=\omega_z/2\pi=42$~Hz, $\kappa/2\pi=2$~MHz, $\gamma/2\pi=0.8$~Hz, $\Delta'=0$, $T_{c,d}=20$~nK,
$l=10$, and $L_p=1$.}
\label{fig:spectrum_sensitivity_sq}
\end{figure}

To obtain the \com{noise spectral density} of the output quadrature, one needs to interfere the cavity output with a local oscillator
to perform homodyne detection.
The injected laser itself can serve as the local oscillator which is combined with the output signal. This allows for the extraction of the phase rotated optical quadrature where the rotation angle $\phi$ 
can be tuned by adjusting a constant phase shift between the two signals. 
We define a generalized quadrature $Q^{\phi}_{\rm out}(\omega)$ at $\phi$ and its \com{spectral density} $S^{\rm{out}}_{Q^\phi,Q^\phi}(\omega)$ as
\begin{align}
Q^{\phi}_{\rm out}(\omega)&= Q_{\rm out}(\omega)\cos\phi + P_{\rm out}(\omega)\sin\phi,\\
S_{Q^{\phi},Q^{\phi}}^{\rm out}(\omega)&=\frac{1}{2\pi}\int_{-\infty}^{\infty}d\omega' \langle Q^{\phi}_{\rm out}(\omega)Q^{\phi}_{\rm out}(\omega')\rangle.
\label{eq:Q_phi_spectra}
\end{align}
By varying the \com{spectral density} over different phase angles, it is noted that $\phi=\pi/2$  is optimal for maximum detection for an unsqueezed input beam i.e., $Q^{\phi}_{\rm out}(\omega)= P_{\rm out}(\omega)$. The optimal \com{spectral density} is then given by
\begin{align}
S^{\rm{{out}}}_{P,P}(\omega)&=A_{P}A_{P}\chi_{QQ} + B_{P}B_{P}\chi_{PP} + A_{P}B_{P}\chi_{QP}\nonumber\\
&+ B_{P}A_{P}\chi_{PQ} + C_{P}C_{P}\chi_{CC} + D_{P}D_{P}\chi_{DD},\nonumber \\
\chi_{XY}&=\frac{1}{2\pi}\int_{-\infty}^{\infty}d\omega'\langle X_{\text{in}}(\omega)Y_{\text{in}}(\omega')\rangle,
\label{eq:spec_gen}
\end{align}
where $X_{\rm in},Y_{\rm in} \in\{Q_{\rm in},P_{\rm in},\epsilon_c,\epsilon_d\}$.
The calculations for the detection angle $\phi$ that optimize the noise spectrum and sensitivity under different conditions are provided in Appendix~\ref{sec:A2}. 

The second important figure of merit is the sensitivity with which the \com{spectral density} and therefore the rotational properties of the BEC can be measured. This is defined as~\cite{Schoenfeld2011,Kumar2021},
\begin{eqnarray}
&\zeta(\omega)=\dfrac{S(\omega)}{\partial S(\omega)/\partial \Lambda}\sqrt{t_{\rm{meas}}},~\zeta_{\rm{opt}}=\zeta(\omega=\omega_{\rm{opt}}),
\label{eq:sensitivity}
\end{eqnarray}
where $t_{\rm meas} \approx \kappa/8a_s^2G^2$ is the measurement time
in the bad cavity regime~\cite{Aspelmeyer2014} and $\omega_{\rm opt}$ is defined as $\left(\partial\zeta/\partial\omega\right)_{\omega=\omega_{\rm{opt}}}=0$. 
The optimal sensitivity $\zeta_{\rm{opt}}$ corresponds to the minima of the function $\zeta(\omega)$ at frequency $\omega_{\rm opt}$. 
The motive here is to find an optimal frequency where the change in spectrum $S(\omega)$ is maximum, which is often relevant in  
experimental implementations using a lock-in detection scheme~\cite{Schoenfeld2011}.
Moreover, $S(\omega)$ responds linearly with 
angular momentum $\Lambda=L_{p}\hbar$ at values of $L_p$ relevant to our analysis, such that  
$S(\omega)\left[\frac{\Delta S(\omega)}{\Delta \Lambda}\right]^{-1}\simeq S(\omega)\left[\frac{\partial S(\omega)}{\partial \Lambda}\right]^{-1}$, which is useful
for numerical estimation of sensitivity.


Figure~\ref{fig:spectrum_sensitivity_sq} shows the variation of $S^{\rm out}_{P,P}(\omega)$ and $\zeta(\omega)$ with frequency $\omega$, 
and studies the effect of introducing squeezed input and atomic collisions in the condensate by solving Eqs.~\eqref{qle1}-\eqref{qle3}. 
The winding number of \com{the input light and the 
BEC} are taken as $l = 10$ and $L_{p}=1$, respectively, which results in peaks of the output spectrum at $\omega_{c~(d)}/2\pi$ at 569 and 695~Hz, respectively, without collisions and  624 and 750~Hz, respectively, with collisions, when $gN = 14$~Hz. 
The optimal sensitivity $\zeta_{\rm opt}$ also occurs at $\omega_{\rm opt}\approx\omega_{c~(d)}$. 
The plots in Fig.~\ref{fig:spectrum_sensitivity_sq} show that the presence of inter-atomic interactions $g$, simply shifts the frequencies of the spectral peaks and also the optimal sensitivity, for both squeezed and unsqueezed light (cf.~\cite{Kumar2021}).
However, for squeezed light with amplitude $r=2$ and angle $\theta=\pi$, the \com{spectral density} drops below the shot noise level, which is an indicator that the noise in the spectrum can be lowered below the quantum \rahul{limit}. For detection angle $\phi = \pi/2$ and squeezing angle $\theta=\pi$, the optimal sensitivity $\zeta_{\rm opt}$ of measurement is not significantly enhanced over the unsqueezed case. 

\begin{figure}[t]
\centering
\includegraphics[width=3in]{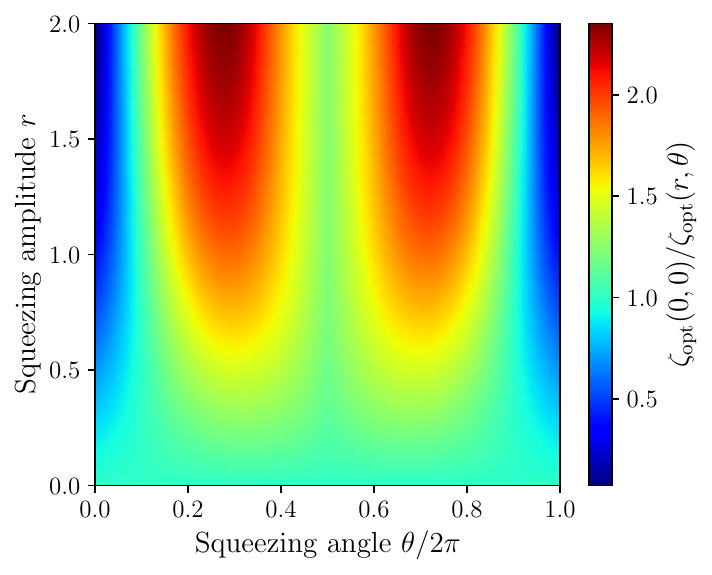}
\caption{Variation of the sensitivity enhancement factor between unsqueezed and squeezed input, as a function of the squeezing amplitude $r$ and angle $\theta$. Here, $\mathcal{P}_{\rm in} = 12.4$~fW.
\com{All the other parameters are the same as in Fig.~\ref{fig:spectrum_sensitivity_sq}}.
}
\label{fig:sensitivity_r_theta}
\end{figure}

%

To better understand the enhancement of measurement sensitivity with squeezing, we look at Fig.~\ref{fig:sensitivity_r_theta}, where the variation of the factor $\zeta_{\rm{opt}}(0,0)/\zeta_{\rm{opt}}(r,\theta)$ with the squeezing amplitude $r$ and angle $\theta$ is shown. For enhancement, the sensitivity with squeezing should be lower, compared to the unsqueezed case, and therefore the factor should be greater than unity.
Notably, $\zeta_{\rm{opt}}(0,0)/\zeta_{\rm{opt}}(r,\theta) \geq 1$  for all values of $r$.
The best enhancement for input power $\mathcal{P}_{\rm in} = 12.4$~fW is observed for angles $\theta=\{2\pi/3,4\pi/3\}$ where a factor of 2 (3 dB) can be achieved, which saturates for $r>2$. 
For $\theta=\pi$, the enhancement factor is close to unity and the sensitivity for squeezed light is similar to those achieved using unsqueezed light, as seen in Fig.~\ref{fig:spectrum_sensitivity_sq}. 
However, the enhancement with squeezing can be much larger at lower values of $\mathcal{P}_{\rm in}$, \com{up to a saturation value around 18 dB}.

%

The optical spectral \com{density}
$S^\text{out}_{P,P}(\omega)$ is given by
%
\begin{align}
&S^{\text{out}}_{P,P}(\omega)=S_{\rm{sn}} + S_{\rm{rp}} + S_{\rm{th}} + S_{\rm{add}},~~{\rm where}\nonumber\\
&S_{\rm{sn}}=B_P(\omega)B_P(-\omega)\frac{\chi_{PP}}{2},\\
&S_{\rm{th}}=C_P(\omega)C_P(-\omega)\chi_{CC} + D_P(\omega)D_P(-\omega)\chi_{DD},\\
&S_{\rm{rp}}=\rm{All~terms~of~}\mathcal{O}\left(G^2\right).
\end{align}
\begin{figure}[t]
\centering
\includegraphics[width=3.2in]{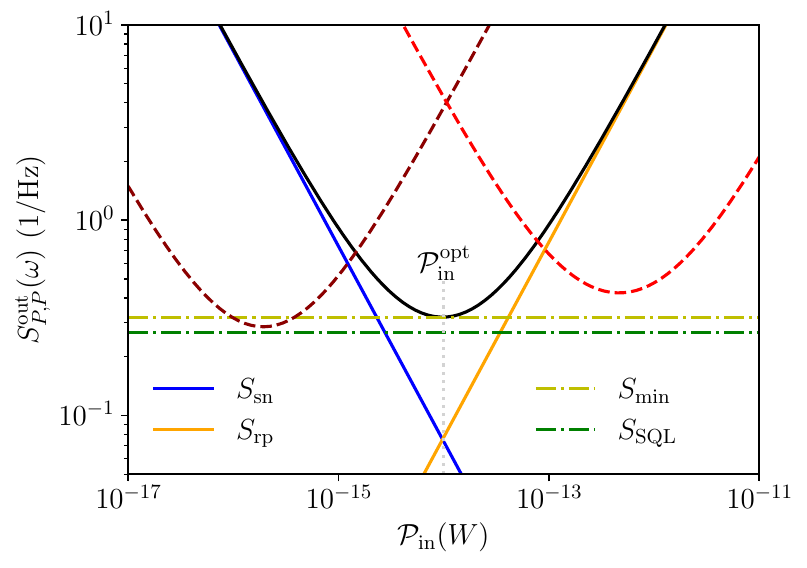}
\caption{The change in optimal output spectrum \com{density} at $\omega_{\rm opt}$ with input power. The output spectrum \com{density} (black-solid) is minimum at an optimal power $\mathcal{P}_{\rm in}^{\rm opt}$, where the decreasing shot noise (blue-solid) and increasing radiation pressure (orange-solid) intersect. The spectrum \com{density} with squeezed input (brown-dashed) with $r=2$, breaks the minimum noise $S_{\rm min}$ (light green-dash-dot) at squeezing angle $\theta=\pi$ at lower input powers. For $\theta=0$, the spectrum density remains above this minimum noise. The spectrum \com{density} remains above the standard quantum limit $S_{\rm SQL}$ (dark green dash-dot) for all squeezing values and input power.
\com{Other parameters are the same as in Fig.~\ref{fig:spectrum_sensitivity_sq}}.
}
\label{fig:sql_approach}
\end{figure}
Note that the spectrum is normalized 
and the additional term $S_{\rm{add}}$ is present only for squeezed input. 
Figure~\ref{fig:sql_approach} shows the variation of the \com{spectral density} $S^{\text{out}}_{P,P}(\omega)$ with input power $\mathcal{P}_{\rm in}$, at the optimal frequency $\omega=\omega_{\rm opt}$. The figure also shows the contribution due to optical shot noise $S_{\rm{sn}}$ and radiation pressure $S_{\rm{rp}}$
\com{(solid blue and orange lines respectively)}.
In the absence of squeezing, the minimum of $S^{\text{out}}_{P,P}(\omega)$ at temperature \com{$T_{c,d}=$20~nK} 
and frequency $\omega_{\rm opt}$ gives us the minimum noise $S_\textrm{min}$. This is achieved at optimal input power $\mathcal{P}_{\rm in}^{\rm out}$, as shown in Fig.~\ref{fig:sql_approach} \com{(solid black curve).}
Now, if $T_{c,d}=0$, the noise $S_{\rm th}$ can be further lowered to ultimately achieve the standard quantum limit $S_\textrm{SQL}$. 
Using squeezed input, with amplitude $r=2$ and $\theta = \pi$, the output \com{spectral density} $S^{\text{out}}_{P,P}$ can be lowered significantly below the minimum noise, thus providing a clear advantage over unsqueezed light for lower values of $\mathcal{P}_{\rm in}$.
However, the minimal \com{noise spectral density} can only get close to SQL but not below it, regardless of the value of $r$ or $\theta$. In fact, for $\theta = 0$, the noise is higher than the minimum noise.
This is primarily due to the measurement backaction noise becoming significant at higher powers. 
As such, using squeezing alone in the input optical light with OAM, the noise cannot be lowered beyond the SQL. \com{To break the SQL one needs to counter the backaction noise using quantum nondemolition measurement~\cite{Thorne1978,Braginsky1980}, which can be achieved with or without using squeezed light~\cite{Kimble2001}.}

%

\section{Rotational sensing using Backaction evasion\label{sec:bae}}

The measurement of a quantum system typically involves its interaction with an additional detection or measuring device. During the measurement
of an observable $\hat{X}$ of the system, system-detector interactions can introduce noise in the system via the variable conjugate to $\hat{X}$, namely $\hat{Y}$, which adversely affects any future measurements of the observable. This noise is called the measurement backaction noise, and was initially evaded by making an effective joint quadrature measurement of the system~\cite{Braginsky1980,Caves1980a,Bocko1996}.
This technique enables the possibility of conducting the quantum non-demolition measurement~\cite{Grangier1998}, where measurement does not affect the observed system. 
The use of backaction evasion (BAE) was used to overcome the SQL in optomechanics~\cite{Clerk2008}, with the technique later being expanded for two-mode measurements~\cite{Tsang2010,Tsang2012}. Further theoretical development has focused on BAE schemes beyond the rotating wave approximation (RWA) to include the effect of counterrotating terms~\cite{Malz2016a,Malz2016b,Brunelli2019}.


For backaction evasion, 
a bichromatic laser beam with two frequencies $\omega_{\mp}=\omega_{a} \pm \delta$ is injected
into the cavity~\cite{Woolley2013} that contains the BEC in a ring trap, where $\delta=\omega_{m}=(\omega_{c} + \omega_{d})/2$ is the average frequency of the two atomic sideband modes.
The optical fields oscillate with amplitudes $\varepsilon_{\pm}$. 
The Hamiltonian is given by,
\begin{align}
&H/\hbar = \omega_a a^{\dagger} a+\omega_c c^{\dagger} c+\omega_d d^{\dagger} d + G a^{\dagger} a\left(X_c+X_d\right)  \nonumber\\
&+ \left(a~\varepsilon_{+}^* e^{i(\omega_a-\omega_m)t}+a~\varepsilon_{-}^* e^{i(\omega_a+\omega_m) t} + {\rm h.c.}\right), 
\label{bae-full-ham}
\end{align}
where we have ignored the terms due to collisions \com{as they only cause a relative shift in the frequencies and have no significant impact on sensitivity.}
Moving to the driving frame\label{eq:driving_transform}
$a = e^{-i \omega_{a} t}\left[\bar{a}\left(e^{-i \delta t}+e^{i \delta t}\right)+\delta a\right]$, where 
$\bar{a}$ and $\delta a$ are the steady state and fluctuations of the cavity field and $\delta$ is the detuning. The linearization of the operator $a$ around a monostable steady state is shown in Appendix~\ref{sec:A3}. An effective Hamiltonian $H_{\rm{eff}}$ is then obtained, which is given by 
\begin{eqnarray}
H_{\text {eff}}/\hbar&=&\left(\omega_{m}+\Omega\right)c^{\dagger}c + \left(\omega_{m}-\Omega\right)d^{\dagger}d\nonumber\\
&+& 2G\bar{a}\cos\delta t\left(X_{c}+X_{d}\right)\left(\delta a^{\dagger} + \delta a\right).
\label{eq:Heff}
\end{eqnarray}
Here, $\Omega = (\omega_c -\omega_d)/2$ and $\delta = \omega_m$ at resonance. 
 %
\begin{figure}[t]
\centering
\includegraphics[width=.8\columnwidth]{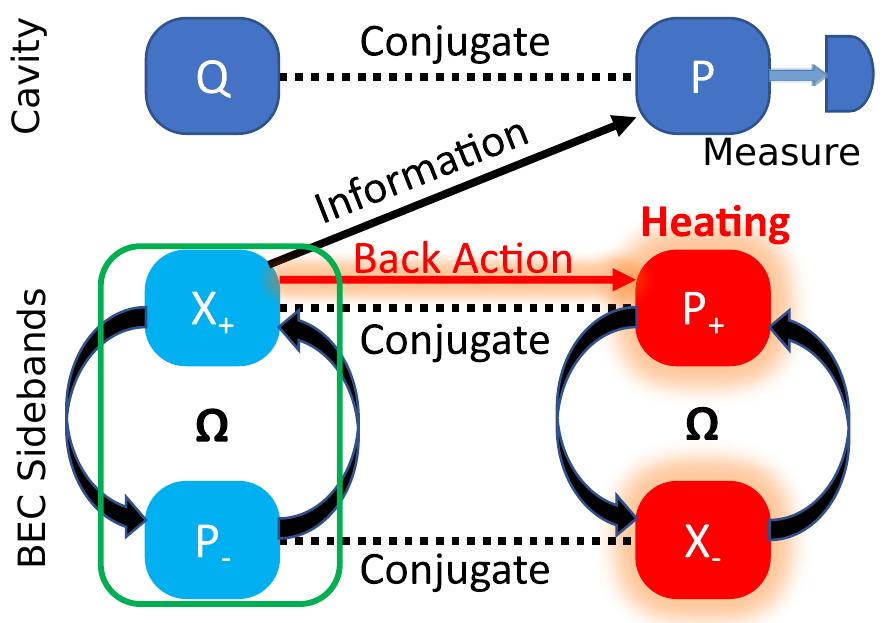}
\caption{A schematic showing the BAE scheme. 
Measurement of the cavity observable $P$ requires information from the $X_{+}$ variable of the BEC. The backaction of the measurement is transferred to the conjugate variable $P_{+}$, which is dynamically decoupled from the subspace of $X_{+}$ and $P_{-}$. Therefore, $P$ can be continuously measured as the evolution of $X_{+}$ is unaffected by backaction.}
\label{fig:bae_scheme}
\end{figure}
In the frame rotating with $\omega_m$, 
the atomic sideband modes oscillate with frequencies $\pm \Omega$, and spectral response will show peaks at these frequencies. The winding number $L_{p}$ of the BEC in the annular trap can be measured using the relation
\begin{equation}
\begin{gathered}
\Omega =\frac{\omega_c - \omega_d}{2} =\frac{\hbar(L_{p} + 2l)^2}{4I} - \frac{\hbar(L_{p} - 2l)^2}{4I}=\frac{2L_{p}l\hbar}{I},
\label{eq:sensitivity_principle}
\end{gathered}
\end{equation}
where $I$ and $l$ are assumed to be known.

The effect of BAE becomes clear under the rotating wave approximation (RWA), where the effective Hamiltonian reduces to a time-independent Hamiltonian, as  $2G\bar{a}\cos(\delta t)=2G\bar{a}$, and under the transformation
$U=e^{i \omega_m (c^\dag c + d^\dag d)t}$, can be written as ($\hbar = 1$)
\begin{equation}
H'_{\text {eff}}=\Omega\left(c^{\dag}c - d^{\dag}d\right) + 2G\bar{a}\cos\delta t\left(X_{c}+X_{d}\right)\left(\delta a^{\dagger} + \delta a\right).\nonumber
\label{eq:Heff_rwa}
\end{equation}
Now, by transforming to a set of symmetric and anti-symmetric quadratures $X_{\pm}=(X_{c} \pm X_{d})/\sqrt{2}, P_{\pm}= (P_{c}\pm P_{d})/\sqrt{2}$, we get
\begin{equation}
H'_{\text {eff}}=\Omega\left(X_+ X_- + P_+ P_-\right)+ \bar{G}X_{+}\delta Q,
\label{eq:symm_ham_rwa}
\end{equation} 
where $\bar{G}=2 \sqrt{2} G \bar{a}$, $\delta Q=(\delta a + \delta a^\dag)/\sqrt{2}$ and $\delta P=i(\delta a^\dag - \delta a)/\sqrt{2}$.
The equation of motion of the system is then given by
\begin{align}
\delta \dot{Q} &= -\frac{\kappa}{2} \delta Q + \sqrt{\kappa} Q_{\text{in}},~~\dot{X}_{\pm}=\Omega P_{\mp},\\
\delta \dot{P} &= -\frac{\kappa}{2} \delta P + \sqrt{\kappa} P_{\text{in}} + \sqrt{2}G\bar{a}X_{+}\delta Q
\label{eq:bae_langevin}
\end{align}
As shown in Fig.~\ref{fig:bae_scheme}, measurement of variable $P$ is dependent on $X_+$, but the measurement backaction on the conjugate variable $P_+$ is decoupled from the dynamics of $X_+$, thus allowing a perfect backaction evasion measurement.

Beyond the RWA, the decoupling of quadratures is not necessarily perfect so as to allow quantum nondemolition measurements, but can still significantly alleviate the effects of backaction noise. 
The key step here is to solve the time-dependent effective Hamiltonian derived in Eq.~\eqref{eq:Heff}. Using the input noise operators and losses from Sec.~\ref{sec:squeezed}, the quantum Langevin equations for the BAE scheme can be written as~\cite{Woolley2013}
\begin{eqnarray}
\delta \dot{a}&=&-\frac{\kappa}{2} \delta a+\sqrt{\kappa} a_{\text{in}}+2 i G \bar{a} \cos (\delta t)\left(X_c+X_d\right) \nonumber \\
&\times& \left(\delta a+\delta a^\dag\right) \label{eq:linearized_langevin_a},\\
\ddot{X}_{c(d)}&=&\left(\omega_m \pm \Omega\right)^2 X_{c(d)}+2 G \bar{a} \cos (\delta t)\left(\omega_m \pm \Omega\right) \nonumber\\
&\times& \left(\delta a+\delta a^\dag \right) -\gamma \dot{X}_{c(d)} +\left(\omega_m \pm \Omega\right) \epsilon_{c(d),\rm in} \label{eq:linearized_langevin_x}.
\end{eqnarray}

\com{In terms of the optical quadrature variables 
$\{Q_{\rm in}, P_{\rm in}\}$, defined in Eq.~\eqref{eq:correlations}},
the above Eq.~\eqref{eq:linearized_langevin_a} can be written as
\begin{align}
\delta \dot{P}= & -\frac{\kappa}{2} \delta P+\sqrt{\kappa}~ P_{\text{in}}+4 G \bar{a} \cos (\delta t) \delta Q\left(X_c+X_d\right), \nonumber\\
\delta \dot{Q}= & -\frac{\kappa}{2} \delta Q+\sqrt{\kappa}~ Q_{\text{in}}.
\label{eq:linearized_langevin_pq}
\end{align}

To obtain the solution for the time-dependent equations of motion
beyond the RWA, we consider an approach~\cite{Malz2016a,Malz2016b} based on the Floquet expansion of the dynamical variables. The formalism is valid even in the bad cavity regime, i.e. for $\omega_{c~(d)} \ll \kappa$ as the couplings are also small $G\ll\kappa$.
Applying the Floquet expansion to each dynamical variable $Z(t)$ as $Z(t)=\sum_{-\infty}^{\infty}Z^{(n)}(t)e^{i n\delta t}$, the variables can be transformed to the frequency domain, by applying the Fourier transform of the above equations. Equating the terms with $e^{i n\delta t}$ and considering the noise terms to be stationary ($n=0$), the following expressions for the quadratures fluctuations is obtained, 
\begin{widetext}
\begin{eqnarray}
\chi_c^{-1}(\omega - n\delta) X_c^{(n)}(\omega)&=&\left[G \bar{a}\left(\delta Q^{(n-1)}+\delta Q^{(n+1)}(\omega)\right)+\epsilon_{c,\text{in}}(\omega)\right.\Bigr] \left(\omega_m+\Omega\right), \label{eq:equations_set1} \\[3pt]
\chi_a^{-1}(\omega-n \delta) \delta Q^{(n)}(\omega)&=&-\delta_{n, 0} \sqrt{\kappa}~ Q_{\rm{in}}(\omega), \label{eq:equations_set2} \\[3pt]
\chi_a^{-1}(\omega-n \delta) \delta P^{(n)}(\omega)&=&-\delta_{n, 0} \sqrt{\kappa}~ P_{\rm{in}}(\omega) + G \bar{a} \sqrt{2}\left(X_c^{(n-1)}+X_c^{(n+1)}+ X_d^{(n-1)} + X_d^{(n+1)}\right), \label{eq:equations_set3} \\[3pt]
\chi_a^{-1}(\omega-n\delta) a^{(n)}(\omega)&=&-\delta_{n, 0} \sqrt{\kappa}~ a_{\mathrm{in}}(\omega)- iG \bar{a}\left(X_c^{(n-1)}+X_c^{(n+1)}+ X_d^{(n-1)}+X_d^{(n+1)}\right)
\label{eq:equations_set4}
\end{eqnarray}
where $\chi^{-1}_{a,c(d)}$ are the susceptibilities defined in Sec.~\ref{sec:squeezed}. Now, the expressions obtained by solving these are 
\begin{eqnarray}
\delta P^{(0)}(\omega) &=& \sqrt{2}\chi^2_a(\omega) G^2 \bar{a}^2 \kappa~ Q_{\rm in}^{(0)} (\omega)\cdot \big[\chi_c(\omega-\delta)+\chi_c(\omega+\delta)+\chi_d(\omega-\delta)+\chi_d(\omega+\delta)\big]\label{eq:full_solution1}, \\[3pt]
\delta P^{(1)}(\omega) &=& \chi_a(\omega-\delta) G \bar{a} \sqrt{2\kappa}~\cdot\big[\chi_c(\omega)\left(\omega_m+\Omega\right) \epsilon_{c,\rm in}^{(0)}(\omega)+\chi_d(\omega)\left(\omega_m-\Omega\right) \epsilon_{d,\rm in}^{(0)}(\omega)\big] \label{eq:full_solution2}, \\[3pt] 
\delta P^{(-1)}(\omega) &=& \chi_a(\omega+\delta) G \bar{a} \sqrt{2\kappa}\cdot\big[\chi_c(\omega)\left(\omega_m+\Omega\right) \epsilon_{c,\rm in}^{(0)}+\chi_d(\omega)\left(\omega_m-\Omega\right) \epsilon_{d,\rm in}^{(0)}(\omega)\big]\label{eq:full_solution3}, \\[3pt] 
\delta P^{(2)}(\omega)&=& \sqrt{2}\chi_a(\omega-2 \delta)G^2\bar{a}^2 \chi_a(\omega)\cdot\big[\chi_{c}(\omega-\delta)+\chi_{d}(\omega-\delta)\big]\sqrt{\kappa}~ Q_{\rm in}^{(0)}(\omega)\label{eq:full_solution4}, \\[3pt] 
\delta P^{(-2)}(\omega)&=& \sqrt{2}\chi_a(\omega+2\delta)G^2\bar{a}^2 \chi_a(\omega)\cdot\big[\chi_{c}(\omega+\delta)+\chi_{d}(\omega+\delta)\big]\sqrt{\kappa}~ Q_{\rm in}^{(0)}(\omega) \label{eq:full_solution5}, \\[3pt] 
\delta P^{(3)}(\omega)&=&0,~~\delta P^{(-3)}(\omega)=0,~{\rm and} ~~\delta Q^{(n)}(\omega) = \delta_{n,0}\sqrt{\kappa}\chi_a(\omega)Q_{\rm in}^{(0)}(\omega).  \label{eq:full_solution6}
\end{eqnarray}
\end{widetext}

\begin{figure}[!h]
\centering
\includegraphics[width=3in]
{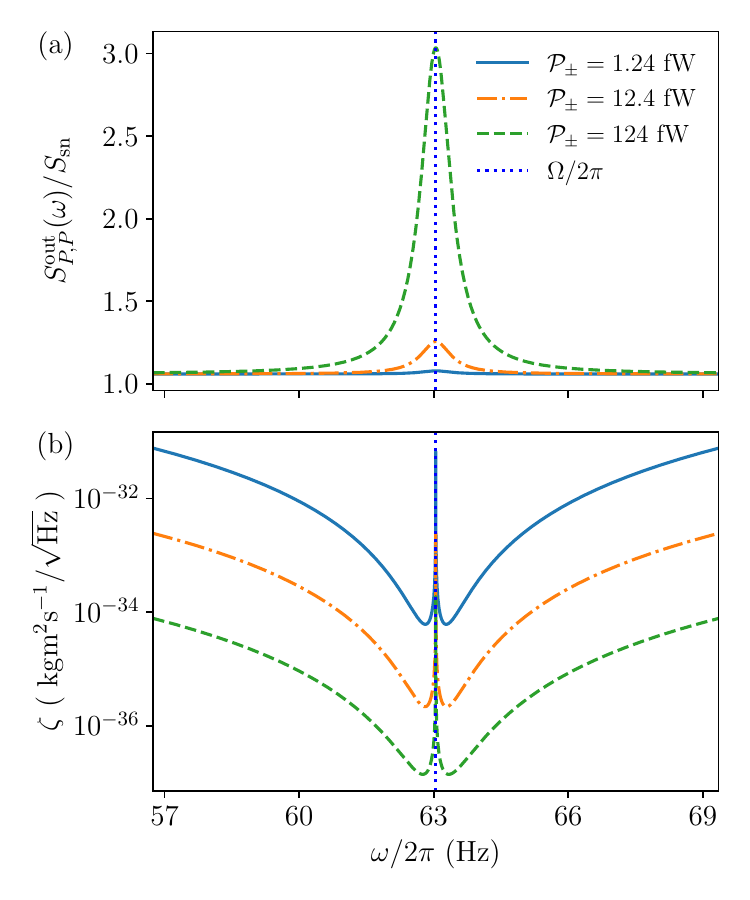}
\caption{Output noise spectrum and measurement sensitivity for the BAE scheme. \com{The plots are for different values of $\mathcal{P}_{\rm in}=\mathcal{P}_\pm$, where $\mathcal{P}_{+}=\mathcal{P}_{-}$ is the same input power for the two driving fields in the BAE scheme.} (a) The plot shows the output optical spectrum $S^\textrm{out}_{P,P}(\omega)$ for different driving frequencies $\omega$, with a peak at frequency $\Omega/2\pi=63$ Hz. \com{The plots have been scaled with the shot noise at different powers for ease of viewing, with $S_\textrm{sn} = 10,1, 0.1$~(in units of 1/Hz) for increasing power.}
b) The variation in sensitivity with $\omega$, with minimal sensitivity occuring at two optimal points $\omega_{\rm opt}$, in the vicinity of the spectral peak. The detection angle is set at $\phi=\pi/2$. \com{Other parameters are the same as in Fig.~\ref{fig:spectrum_sensitivity_sq}}.
}
\label{fig:spectrum_sensitivity}
\end{figure}

Importantly, the fact that the input quantum and thermal noise is stationary i.e., only the zeroth Floquet mode ($n=0$) is nonzero, ensures that the series truncates at $|n|=3$ for all noise quadratures. 
\com{From Eq.~\eqref{eq:equations_set2}, it is noted that $\delta Q^{(n)}$ is nonzero only for $n=0$, which implies that modes in quadrature $\delta X_c^{(n)}$ in Eq.~\eqref{eq:equations_set1} are $n=0,\pm 1$. Hence, operators $\delta P^{(n)}$ and $a^{(n)}$ in Eqs.~\eqref{eq:equations_set3}- Eq.~\eqref{eq:equations_set4} have Floquet modes only for $n=0,\pm 1,\pm 2$.
As such for stationary noise, 
the optical susceptibilities for frequency $\omega - n\delta$ can at most add up to the  second harmonics (0, $\omega \pm \delta$, $\omega \pm 2\delta$), giving rise to contribution from only a finite set of sideband modes.}
This reduces the complexity in solving the problem and in fact allows for exact solutions to be obtained for the noise quadrature fluctuations. 
The nonzero Floquet components are then used to calculate the zeroth order \com{spectral density}, which represents the average optical spectrum. The spectral response function will allow us to directly estimate the rotational  properties of the BEC.


\begin{figure*}[t]
\centering
\includegraphics[width=6in]{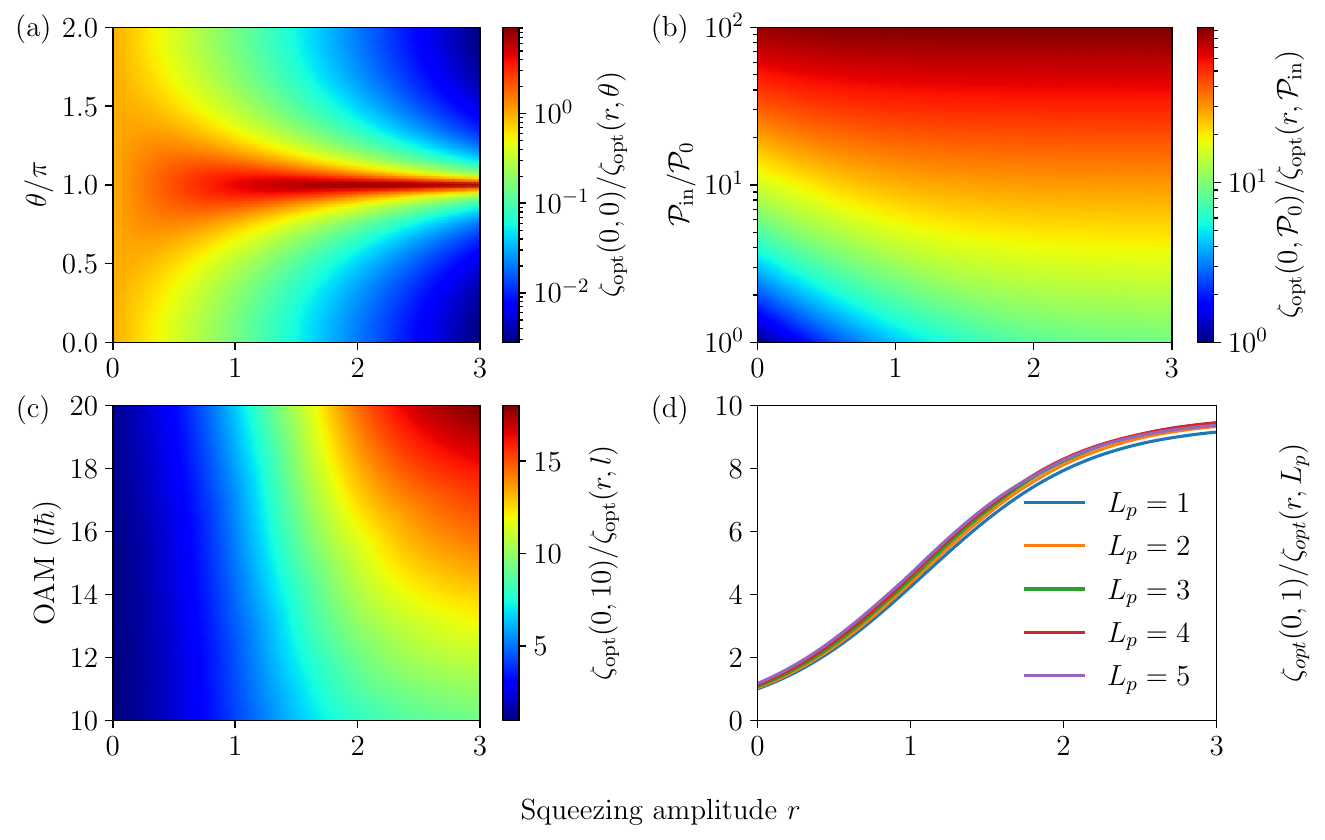}
\caption{Enhancement of optimal sensitivity in the BAE scheme using squeezed input light. The plot shows the enhancement as a function of squeezing amplitude $r$ and a) squeezing angle $\theta$,  b) relative input power $\mathcal{P}_{\rm in}/\mathcal{P}_0$, c) \ncom{OAM of the input light $l\hbar$}, and d) \ncom{the winding number of the atomic condensate $L_p$}. 
The input power $\mathcal{P}_{\rm in}=\mathcal{P}_{0} = 12.4$~fW and $l = 10$, unless varied. 
\com{All the other parameters are the same as in Fig.~\ref{fig:spectrum_sensitivity_sq}}.
}
\label{fig:squeezed_bae}
\end{figure*}

Now to obtain the output spectrum, the input-output formalism is again used. 
Let $\{A(\omega), B(\omega), C(\omega), D(\omega)\}$ be the frequency-dependent coefficients of the noise quadrature operators $Q_{\text{in}},P_{\text{in}},\epsilon_{c,\text{in}}$ and $\epsilon_{d,\text{in}}$, 
respectively in Eqs.~\eqref{eq:full_solution1}-\eqref{eq:full_solution6}. The expressions for these coefficients are shown in Appendix~\ref{coeff_bae}. 
The homodyne detection angle is set at $\phi = \pi/2$, as the noise spectrum is analysed using the noise quadrature $P_{\rm out}(\omega)$, which is given by

\begin{eqnarray}
P^{(n)}_{\rm out}(\omega) &=& A^{(n)}(\omega)Q_{\rm in}(\omega) + B^{(n)}(\omega)Q_{\rm in}(\omega)\nonumber\\
&+& C^{(n)}(\omega)\epsilon_{c,\rm in}(\omega) + D^{(n)}(\omega)\epsilon_{d,\rm in}
\label{eq:P_decompose}
\end{eqnarray}
The Fourier components of the spectrum~\cite{Malz2016a} are given by the expression
\begin{equation}
S_{A^{\dagger} A}^{(m)}(\omega)=\sum_n \int \frac{d \omega^{\prime}}{2 \pi}\langle A^{(n) \dagger}(\omega+n \delta) A^{(m-n)}\left(\omega^{\prime}\right)\rangle,
\label{eq:general_spectrum}
\end{equation}
where the time-averaged noise spectrum is captured by the zeroth Fourier component ($n=0$) of the above spectrum, given by
\begin{equation}
S_{A^{\dagger} A}^{(0)}(\omega)=\sum_n \int \frac{d \omega^{\prime}}{2 \pi}\left\langle A^{(n) \dagger}(\omega+n \delta) A^{(-n)}\left(\omega^{\prime}\right)\right\rangle.
\end{equation}
Using Eqs.~\eqref{eq:full_solution1}-\eqref{eq:full_solution6} and the coefficients in  Eq.~\eqref{eq:P_decompose}, the expression for the \com{spectral density} is
\begin{align}
S_{P,P}^{\rm out}(\omega)&=\frac{1}{2}\left[A^{-2}(\omega+2 \delta) A^{+2}(-\omega-2 \delta) \chi_{QQ}(\omega+2\delta)\right. \nonumber\\
&\left.+ C^{-1}(\omega+\delta) C_1(\omega-\delta) \chi_{cc}(\omega+\delta)\right.\nonumber\\
&+D^{-1}(\omega+\delta) D_1(-\omega-\delta) \chi_{dd}(\omega+\delta)\nonumber\\
&+A^{2}(\omega-2 \delta) A^{-2}(-\omega+2 \delta) \chi_{QQ}(\omega-2\delta) \nonumber\\
&+ C^{1}(\omega-\delta) C_{-1}(\omega+\delta) \chi_{cc}(\omega-\delta)\nonumber\\
&\left. +D^{1}(\omega-\delta) D^{-1}(-\omega+\delta) \chi_{dd}(\omega-\delta)\right]\nonumber\\
& + \frac{1}{2}\left[A^0(\omega)A^0(-\omega) \chi_{QQ}(\omega) + B^0(\omega)B^0(-\omega)\chi_{PP}\right.\nonumber\\
&\left.+ A^0(\omega)B^0(-\omega)\chi_{QP} + B^0(\omega)A^0(-\omega)\chi_{PQ}\right]
\label{eq:spectrum_bae}
\end{align}
where $\chi_{i,j}(\omega)$ are the input noise correlations between $i$ and $j$ modes such that $i,j\in\{c_{\rm in}(\omega),d_{\rm in}(\omega),Q_{\rm in}(\omega),P_{\rm in}(\omega)\}$. Figure~\ref{fig:spectrum_sensitivity}(a) shows the output spectral 
function $S_{P,P}^{\rm out}(\omega)$ \com{for different values of the 
input power $\mathcal{P}_\textrm{in} = \mathcal{P}_\pm$},
with the peak occurring at $\omega=\Omega$. 
%
The peak is directly related to the winding number $L_{p}$ of the BEC and is sharper for \com{higher input power imbalance.}
The measurement sensitivity is given by $\zeta(\omega)$ defined in Eq.~\eqref{eq:sensitivity}, where the optomechanical measurement time is given by 
$t^{-1}_{\rm{meas}}(\omega) = \big|{\sqrt{2\kappa}G\bar{a}}/{\left(-i\omega + \kappa /2\right)}\big|^2$. This is defined as the absolute value of the rate at which the output quadrature mode $P^{(0)}_{\rm out}$ changes with respect to the mechanical fluctuations $\delta X_{c(d)}^{(\pm 1)}$. In Fig.~\ref{fig:spectrum_sensitivity}(b), the sensitivity $\zeta$ is varied with frequency $\omega$, for different input powers $\mathcal{P}_\pm$. Similar to the squeezed input case, the maximum sensitivity occurs at two points in the close vicinity of the spectral peak $\Omega$. The optimal frequency $\omega_{\rm opt}$ can then be used to precisely sense the angular momentum of the rotating BEC inside the cavity.

\begin{figure}[t]
\centering
\includegraphics[width=3.1in]{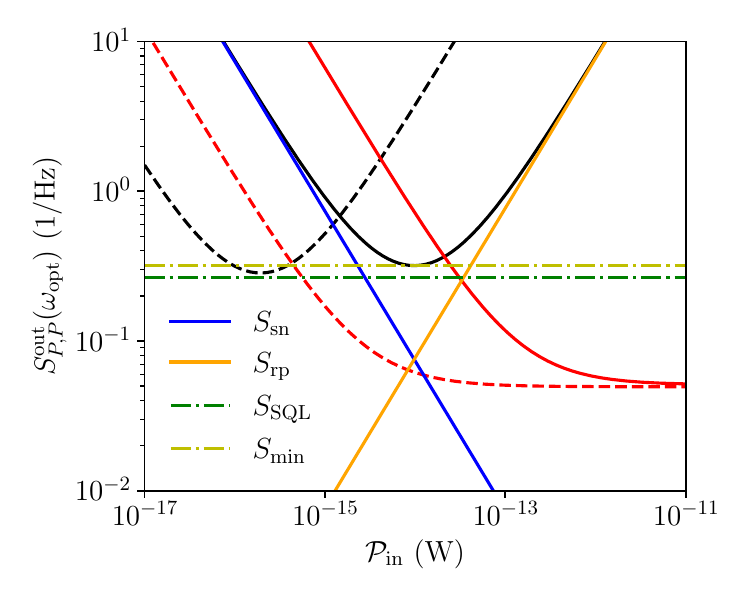}
\caption{The output spectrum with input power for the BAE scheme. The output spectrum (black-solid curve) due to the contributions from shot noise (blue-solid line), radiation pressure (orange-solid line), and thermal and quantum fluctuations, which reaches minimum value $S_{\rm min}$ (green-dash-dot line) at input power $\mathcal{P}_{\rm in}^{\rm opt}$.
The spectrum due to the backaction evasion scheme (red-solid) is significantly lower than the standard quantum limit (light green dash-dot line) at higher input powers. For squeezed input in bichromatic BAE driving, with $r=2,\theta=\pi$, the spectrum can break the SQL even at lower $\mathcal{P}_{\rm in}$ (red-dashed curve). \com{All the other parameters are the same as in Fig.~\ref{fig:spectrum_sensitivity_sq}}.
}
\label{fig:sql_break}
\end{figure}

The sensitivity of the \com{backaction evasion measurements} can be enhanced and the output noise lowered by adding squeezing to the bichromatic drive~\cite{Ghosh2022}. 

As in Sec.~\ref{sec:squeezed}, in this case the squeezing is introduced through the input noise correlations, with amplitude $r$ and angle $\theta$. The enhancement of sensitivity with squeezed light in the BAE method is shown in Fig.~\ref{fig:squeezed_bae}, for different parameter regimes.
In all cases, the enhancement factor is the ratio of the optimal sensitivity, \com{with and without squeezed light in the backaction evasion approach, varied along with some other parameter such as power or OAM of the input light}.

For instance in Fig.~\ref{fig:squeezed_bae}(a), the enhancement factor is $\zeta_{\rm opt}(0,0)/\zeta_{\rm opt}(r,\theta)$, which shows the variation of enhancement with $r$ and $\theta$. It is observed that $\theta = \pi$ always generates the maximum sensitivity and is thus fixed for all the other plots. In Fig.~\ref{fig:squeezed_bae}(b), the variation of the enhancement factor with the input power $\mathcal{P}_{\rm in}$ and squeezing amplitude $r$ is shown. At low squeezing, the sensitivity decreases with higher power and the enhancement factor can increase by nearly two orders of magnitude. Importantly, at high $\mathcal{P}_{\rm in}$, where the radiation pressure noise is dominant, the effect of squeezing becomes marginal.

The variation of enhancement factor with OAM number $l$ of the input light is shown in Fig.~\ref{fig:squeezed_bae}(c), where it is evident that enhancement is higher for larger values of $l$ and $r$.  
\com{Finally, the enhancement in sensitivity is shown to only marginally change for different values of the angular momentum of the condensate $\ncom{L_p\hbar}$, as shown in Fig.~\ref{fig:squeezed_bae}(d).}

\com{The variation of the output \ncom{noise spectral density} $S_{P,P}^{\rm out}(\omega)$ at the optimal frequency $\omega=\omega_{\rm opt}$ with the input power $\mathcal{P}_{\rm in}$ is shown in Fig.~\ref{fig:sql_break}.}
While the \ncom{spectral density} could only be lowered below the minimum noise at temperature $T_{c,d}$ 
using squeezed light (see Fig.~\ref{fig:sql_approach}), using the backaction evasion method the noise can be lowered well beyond the standard quantum limit (SQL). 
\com{This is achieved by significantly lowering the radiation pressure noise through the backaction evasion method, which is typically more prominent at higher input powers. In contrast, at lower input powers, where the shot noise is more dominant, the noise spectrum is not significantly lowered. However, by introducing squeezing in the input bichromatic drive of the BAE scheme, the shot noise barrier can be broken and noise beyond the SQL can be achieved even at lower input powers.} 

\begin{figure}[h]
\centering
\includegraphics[width=3.0in]{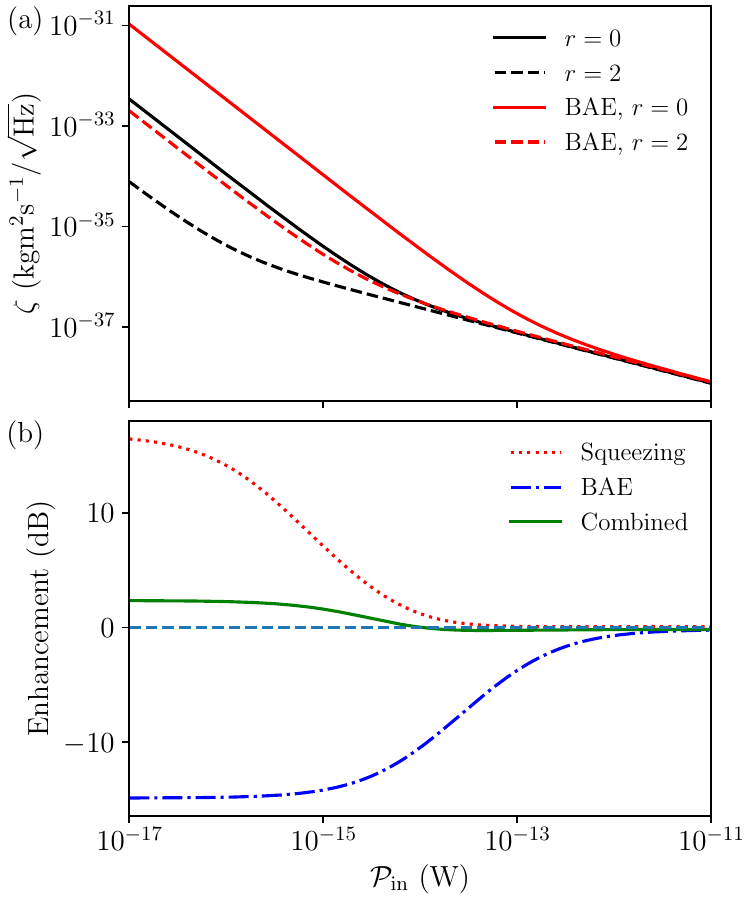}
\caption{Sensitivity and its enhancement for different values of input power $\mathcal{P}_{\rm in}$. (a) The plot shows the sensitivity with (red-solid) and without (black-solid) the BAE approach. The effect of squeezing in each case is shown by the dashed line. (b) The plot \com{shows the enhancement in} sensitivity due to squeezed input alone (red-dotted), BAE approach (blue-dash-dot), and BAE with additional squeezing (green-solid), in comparison to driving the cavity with only OAM light. For all input $l = 10$, and for squeezed noise, $r=2$ and $\theta=\pi$.   \com{All the other parameters are the same as in Fig.~\ref{fig:spectrum_sensitivity_sq}}.
}
\label{fig:sense_comp}
\end{figure}

\com{An important result in the detection of rotational properties of the BEC is the trade-off involved in achieving enhanced sensitivity using squeezed light and on the other hand, lowering the \ncom{noise spectral density} far beyond the SQL using nondemolition measurements such as the backaction evasion method. This is evident from the sensitivity enhancement for the two different protocols, as observed in Fig.~\ref{fig:sense_comp} 
for a wide range of input powers $\mathcal{P}_{\rm in}$.} 
\com{The enhancement is in comparison to the initial case where a monochromatic, unsqueezed OAM beam is injected to the cavity and interacts with the condensate.
At lower input powers, where shot noise is dominant, the effect of using squeezing in the OAM of light on sensitivity of measurements is quite strong and an enhancement of 18~dB can be observed. However, for input powers above $\mathcal{P}_\textrm{in}=\mathcal{P}_0$, there is no enhancement in sensitivity. The use of backaction evasion to lower the noise beyond SQL, seems to have the opposite effect on measurement sensitivity, especially at lower powers. The bichromatic drive in BAE method broadens the \ncom{noise spectral density} and makes it significantly less sensitive. However, if squeezing is added to the BAE scheme, sensitivity enhancement of 2-3~dB is observed at low powers. Importantly, at $\mathcal{P}_\textrm{in}=\mathcal{P}_0$ the sensitivity is as good as the non-BAE approaches, and hence the noise can lowered without negatively affecting the sensitivity of the measurements.
At powers above $\mathcal{P}_{\textrm{in}}\sim 1$ pW, none of the approaches provide any enhancement over the initial case.}

\section{Conclusion\label{sec:conc}}
In this work we have \com{analyzed} the measurement of the rotation of an atomic BEC in a ring
trap, such as its winding number, using cavity modes carrying orbital angular momentum, in a minimally destructive manner. The interaction between the sidemodes of the BEC and the 
light carrying OAM
reduces to a set of optomechanical equations of motion, which typically allows for sensing of the angular momentum close to the standard quantum limit.

By exploiting the theoretical framework of optomechanics, we have shown that the quantum sensing of such rotational properties can be significantly enhanced. The introduction of squeezed light as input, and for optimal homodyne detection angles, the sensitivity of detecting the spectral peaks of the output light and therefore the angular momentum, can be enhanced by a factor of $2-100$  ($\approx 4-20~{\rm dB}$) as compared to unsqueezed input. Notably, the higher enhancement occurs only at low input powers, \com{where shot noise is dominant}. Moreover, the optical quadrature noise, at optimal frequencies $\omega$, is observed to go below the shot noise and break the minimum noise level at temperature $T_{c,d}$, which was hitherto the best that could have been achieved using OAM light as input, without any squeezing.

\com{The noise spectral density can be reduced below the SQL by 
using backaction evasion methods to counter the radiation pressure or 
measurement backaction noise at high input powers}. This was achieved by using a bichromatic drive, which yields a spectral peak at a frequency directly related to the winding number of the BEC. 
Using an approach based on Floquet theory to solve the time-dependent dynamics, the output \com{noise spectral density} was calculated in the bad cavity regime, and beyond the rotating wave approximation. 
\com{It was observed that the backaction evasion allowed the output \com{noise spectral density} to be lowered beyond the standard quantum limit, albeit at the cost of measurement sensitivity.}
\com{However, using squeezed light in the bichromatic drive of the backaction evasion method, the overall measurement sensitivity was improved, and at relevant input powers, the sensitivity is restored to non-BAE levels with SQL being broken even at lower input powers.
Ultimately, the use of squeezing in the BAE scheme 
allowed for very precise measurements of rotational properties of the BEC, with noise well below the standard quantum limit.}


The framework introduced in the study is quite extensive and can be applied to investigate other properties and phenomena related to the study of atomic superfluids and its interaction with nonclassical light. We expect our results to be relevant to 
ongoing studies of superfluid hydrodynamics, and the sensing and manipulation of rotating matter waves. Of further interest is the role of ponderomotive squeezing introduced in the scattered atomic sidemodes and the resulting entanglement in the system. Moreover, full quantum solutions in systems with low excitations can allow for applications that are directly related to quantum technology such as generation of entangled light, coherent transfer of information between the collective states of light and matter and implementation of lower power quantum devices.

\begin{acknowledgments}
RG acknowledges the financial support from CSIR-HRDG, India in the form of SRF. P.K. acknowledges financial support from the Max Planck Society. MB would like to thank the Air Force Office of Scientific Research (FA9550-23-1-0259) for support. R.K. acknowledges support from JSPS KAKENHI Grant No. JP21K03421. HSD acknowledges support from SERB-DST, India under a Core-Research Grant (No: CRG/2021/008918) and from IRCC, IIT Bombay (No: RD/0521-IRCCSH0-001).
\end{acknowledgments}

\appendix

\section{Scattering of BEC with optical lattice formed by OAM beams\label{sec:A0}}
As stated in Sec.~\ref{sec:model}, the cavity is injected with a single mode, symmetric superposition of OAM light given by 
$\ket{+}=\ket{+l\hbar}+\ket{-l\hbar}$ 
(with annihilation operator $a$). 
The light-matter interaction 
varies azimuthally as
$\sim\cos^2(l\varphi)a^\dag a$. This changes the spatial mode of BEC from $L_p$ as follows:
\begin{eqnarray}
\cos ^2(l \varphi)\left|L_p\right\rangle&=&\left[\frac{1+\cos (2 l \varphi)}{2}\right]\left|L_p\right\rangle, \nonumber \\
&=& \frac{1}{2}\left|L_p\right\rangle+\frac{1}{4}\left(e^{i 2 l \varphi}+e^{-i 2 l \varphi}\right)\left|L_p\right\rangle, \nonumber \\
&=& \frac{1}{2}\left|L_p\right\rangle+\frac{1}{4}\left(\left|L_p+2 l\right\rangle+\left|L_p-2 l\right\rangle\right). \nonumber \\
\end{eqnarray}
In the last line, we have used the fact that the exponential of the angular displacement operator $\varphi$ is the generator of angular
momentum translations. Now the BEC order parameter in the position representation is given by
\begin{align}
\Phi(\varphi) &\sim\left\langle\varphi\left|\cos ^2(l \varphi)\right| L_p\right\rangle \sim \frac{1}{2}\left\langle\varphi \mid L_p\right\rangle\nonumber\\
&+\frac{1}{4}\left(\left\langle\varphi \mid L_p+2 l\right\rangle+\left\langle\varphi \mid L_p-2 l\right\rangle\right) \nonumber \\
& \sim \frac{e^{i L_p \varphi}}{2}+\frac{e^{i\left(L_p+2 l\right) \varphi}+e^{i\left(L_p-2 l\right) \varphi}}{4}\nonumber\\
 &\sim \frac{e^{i L_p \varphi}}{2}\left[1+\cos (2 l \varphi)\right].
\end{align}
The density is given by $n(\varphi)\sim\vert\Phi(\varphi)\vert^2\sim\left[1+ \cos (2 l \varphi)\right]^2$, which shows that only the $\cos(2l\varphi)$ and $\cos(4l\varphi)$ density modes are populated by scattering from the $\ket{+}$ mode.

Importantly, sine density modes, which correspond to 
light-matter interaction arising from
antisymmetric superposition of OAM light
$\ket{-}=\ket{+l\hbar}-\ket{-l\hbar}$, are not present if only first-order Bragg scattering processes are considered. In fact, since higher-order scattering terms involve repetitive action of $\cos^2(l\varphi)$ on the scattered states, these processes also do not introduce sine mode density excitations. 
However, interference between the persistent currents can introduce sine modes. Nonetheless, these condensate excitations cannot transfer photons out from
the $\ket{+}$ mode to any other optical mode, since Bragg scattering involves stimulated emission, which can repopulate only the $\ket{+}$ mode~\cite{Stenger1999}. Photons can be scattered into modes other than $\ket{+}$ only via spontaneous emission. However, for the large light-atom detunings considered in this work, spontaneous emission is negligible \cite{Ferdinand2008,Kumar2021}. Thus, we only consider the $\ket{+}$ optical mode in our model.

\section{Coefficients of output quadratures\label{sec:A1}}
To obtain the output optical quadrature $\delta P(\omega)$ and $\delta Q(\omega)$, the equation $\mathcal{F}\delta u = D u_{\rm in}$ needs to be solved, which requires inverting the matrix $\mathcal{F}$ in Eq.~\eqref{eq:matrix}. Solving the equation, and following the input-output formalism, the following coefficients for the quadratures are obtained
\begin{figure*}[t]
\centering
\includegraphics[width=7.2in]{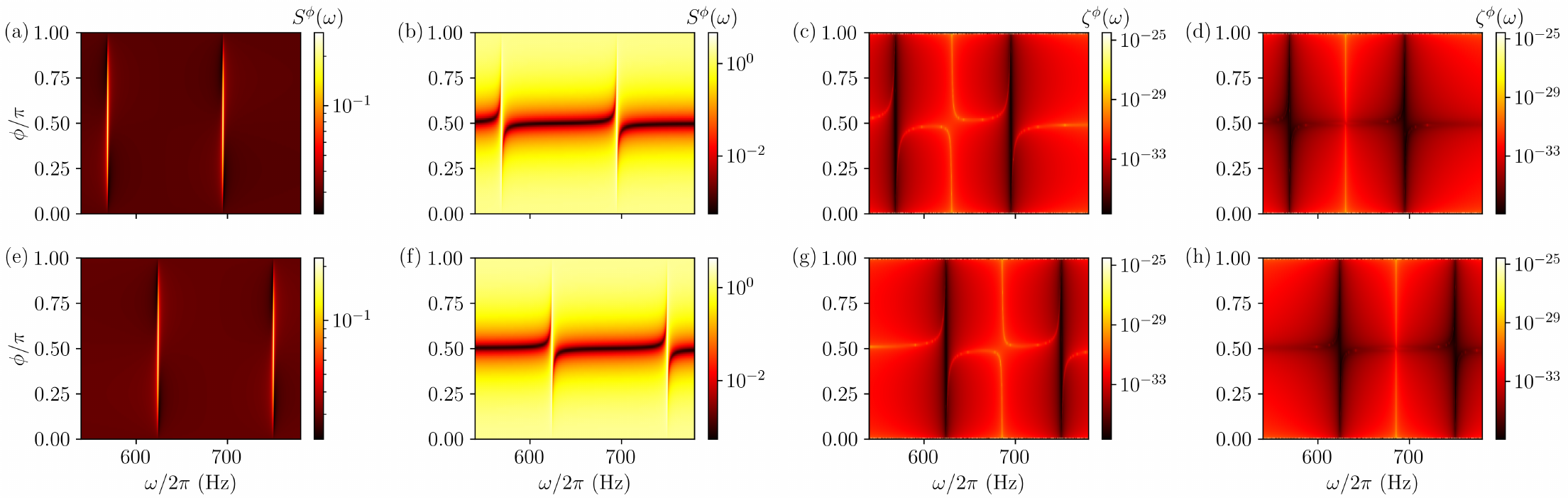}
\caption{Variation of output optical spectrum and sensitivity with detection angle $\phi$. 
The plots in (a)-(e) and (c)-(g), are for output spectrum and sensitivity, respectively, for unsqueezed input light,
while (b)-(f) and (d)-(h) are the same, but for squeezed input with amplitude $r=2$ and angle $\theta = \pi$. 
The plots in the top (bottom) row correspond to cases with (without) inter-atomic interaction or collisions.
}
\label{fig:colormap_phi_var}
\end{figure*}
\begin{eqnarray}
P_{\rm out}(\omega)&=& A_{P}(\omega)Q_{\rm in}(\omega) + B_{P}(\omega)Q_{\rm in}(\omega)\nonumber\\ 
&+& C_{P}(\omega)\epsilon_{c,\rm in}(\omega) + D_{P}(\omega)\epsilon_{d,\rm in},\\[3pt]
Q_{\rm out}(\omega)&=& A_{Q}(\omega)Q_{\rm in}(\omega) + B_{Q}(\omega)P_{\rm in}(\omega) \nonumber\\
&+& C_{Q}(\omega)\epsilon_{c,\rm in}(\omega) + D_{Q}(\omega)\epsilon_{d,\rm in},
\label{eq:PQ_decompose_sq}
\end{eqnarray}
where $\{A_{P(Q)},B_{P(Q)},C_{P(Q)},D_{P(Q)}\}$ are the output optical quadrature coefficients for respective components of 
$\{Q_{\rm in},P_{\rm in},\epsilon_c,\epsilon_d\}$. 
The coefficients satisfy, 
$A_{P}/A'_{P} = C_{P}/C'_{P} = D_{P}/D'_{P}, B_{Q}/B'_{Q}=C_{Q}/C'_{Q}=D_{Q}/D'_{Q}$ = $\sqrt{\kappa}$, and $A_{Q}/A'_{Q}=B_{P}/B'_{P}=\sqrt{\kappa}-1$, where
\begin{eqnarray}
A'_Q(\omega)&=&{\chi _a \sqrt{\gamma _o} \left(\mathcal{A}^2 \chi_c \chi_d+1\right)}/{\mathcal{D}(\omega)},\\[3pt]
B'_Q(\omega)&=&-{\Delta'  \chi _a^2 \sqrt{\gamma _o} \left(\mathcal{A}^2 \chi_c \chi_d+1\right)}/{\mathcal{D}(\omega)},\\[3pt]
C'_Q(\omega)&=&{\sqrt{2} \Delta'  G a_s \chi _a^2 \chi_c \omega _c \left(\mathcal{A} \chi_d+1\right)}/{\mathcal{D}(\omega)},\\[3pt]
D'_Q (\omega)&=&{\sqrt{2} \Delta'  G a_s \chi _a^2 \chi_d \omega _d \left(1-\mathcal{A} \chi_c\right)}/{\mathcal{D}(\omega)},\\[3pt]
A'_P (\omega)&=&\left[\sqrt{\gamma _o}\chi _a^2 \left(\chi_c \left(2 G^2 a_s^2 \tilde{\omega }_c \left(\mathcal{A} \chi_d+1\right)+\mathcal{A}^2 \Delta' \chi_d\right)\right.\right.\nonumber\\[3pt]
&-& 2 \left.\left.G^2 a_s^2 \chi_d \tilde{\omega}_d \left(\mathcal{A} \chi_c-1\right)+\Delta' \right)\right]/{\mathcal{D}(\omega)},\\[3pt]
B'_P (\omega) &=& {\chi _a \sqrt{\gamma _o} \left(\mathcal{A}^2 \chi_c \chi_d+1\right)}/{\mathcal{D}(\omega)},\\[3pt]
C'_P (\omega) &=& -{\sqrt{2} G a_s \chi _a \chi_c \omega _c \left(\mathcal{A} \chi_d+1\right)}/{\mathcal{D}(\omega)},\\[3pt]
D'_P (\omega) &=& {\sqrt{2} G a_s \chi _a \chi_d \omega _d \left(\mathcal{A} \chi_c-1\right)}/{\mathcal{D}(\omega)},
\\[3pt]
\mathcal{D}(\omega) &=& \Delta'  \chi _a^2 \left(\chi_c \left(2 G^2 a_s^2 \tilde{\omega }_c \left(\mathcal{A} \chi_d+1\right)+\mathcal{A}^2 \Delta' \chi_d \right) \right. \nonumber\\
&-& \left. 2 G^2 a_s^2 \chi_d \tilde{\omega }_d \left(\mathcal{A} \chi_c-1\right)+\Delta' \right) + \mathcal{A}^2 \chi_c \chi_d+1. 
\nonumber\\
\end{eqnarray}

\section{Optimal homodyne detection angle for noise spectrum and  sensitivity\label{sec:A2}}

For the homodyne detection, one can define generalized quadratures with respect to an angle $\phi$. The output \com{spectral density} for the generalized quadrature is given by Eq.~\eqref{eq:Q_phi_spectra} and the different components are given by
$
S_{Q^\phi, Q^\phi}^{\rm out}=S^\phi(\omega)=\cos ^2 \phi S_{Q Q}+\sin ^2 \phi S_{P P}+\cos \phi \sin \phi\left(S_{Q P}+S_{P Q}\right).
$
To maximize this \com{spectral density}, ${\partial S^\phi}/{\partial \phi}=0$, with ${\partial^2 S^\phi}/{\partial \phi^2}<0$, which gives 
\begin{eqnarray}
\frac{\partial S^\phi}{\partial \phi}&=&\sin 2 \phi\left(S_{P P}-S_{Q Q}\right)+\cos 2 \phi\left(S_{Q P}+S_{P Q}\right), \nonumber\\
&=& 0, ~ \Rightarrow \phi=\frac{1}{2} \tan ^{-1}\left(\frac{S_{Q P}+S_{P Q}}{S_{Q Q}-S_{P P}}\right). \label{max_cond_1}\\
\frac{\partial^2 S^\phi}{\partial \phi^2} &=& 2 \cos 2 \phi\left(S_{P P}-S_{Q Q}-\tan 2 \phi\left(S_{Q P}+S_{P Q}\right)\right), \nonumber\\
&<& 0,~ \Rightarrow \quad \frac{S_{P P}-S_{Q Q}}{S_{Q P}+S_{P Q}}<\tan 2 \phi. \label{max_cond_2}
\end{eqnarray}
Using relations Eqs.~\eqref{max_cond_1} and \eqref{max_cond_2},
\begin{align}
\frac{S_{P P}-S_{Q Q}}{S_{Q P}+S_{P Q}}&<\frac{S_{Q P}+S_{P Q}}{S_{Q Q}-S_{P P}}, \\ 
\Rightarrow -\left(S_{P P}-S_{Q Q}\right)^2&<\left(S_{Q P}+S_{P Q}\right)^2, \nonumber\\ 
\Rightarrow \left(S_{Q P}+S_{P Q}\right)^2 &+\left(S_{P P}-S_{Q Q}\right)^2 > 0.
\end{align}
%
\begin{align}
S^{\phi}(\omega)&=\frac{1}{2}\left(S_{Q Q}+S_{P P}\right)\nonumber\\
&+\frac{1}{2} \sqrt{\left(S_{Q Q}-S_{P P}\right)^2+\left(S_{Q P}+S_{P Q}\right)^2}.
\end{align}
For unsqueezed input, $S_{Q P}=-S_{P Q}$ and $S_{P P}>S_{Q Q}$, which gives the detection angle $\phi_0=\pi/2$ for the optimal noise spectrum. Therefore, $Q_{\rm out}^\phi = P_{\rm out}$ and $S^\phi=S^{\rm out}_{P,P}$.
However, the above conditions for optimal $\phi$ may vary if squeezed input is used. 

\begin{figure}[t]
\centering
\includegraphics[width=\columnwidth]{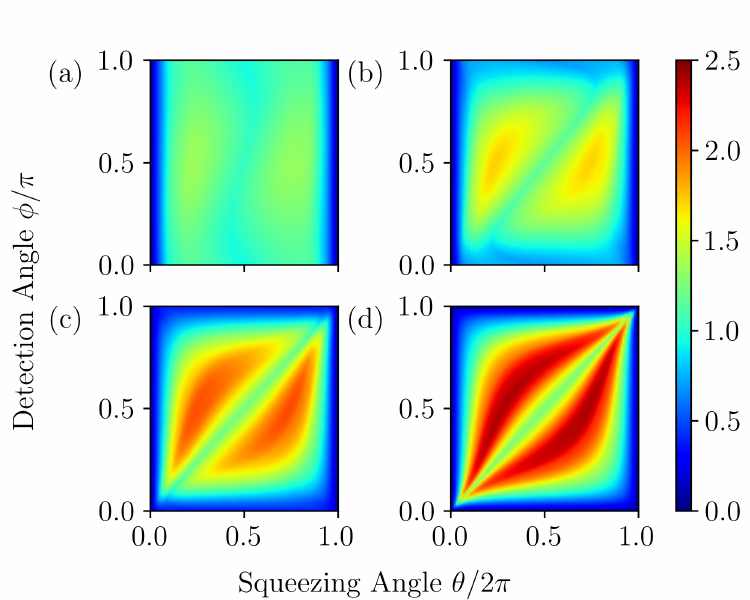}
\caption{Variation of sensitivity enhancement factor with detection angle $\phi$ and squeezing angle $\theta$, for different values of squeezing amplitude, $r = 0.1$ (top left), $r = 0.5$ (top right), $r = 1.0$ (bottom left) and $r = 2.0$ (bottom right).} 
\label{fig:zeta(phi_theta)}
\end{figure}

Figure~\ref{fig:colormap_phi_var} shows the numerical results for the \com{spectral density} $S^{\phi}(\omega)$ and the sensitivity $\zeta^{\phi}(\omega)$, obtained by solving the quantum Langevin equations, as a function of the homodyne detection angle $\phi$. The figures consider the optimal spectrum and sensitivity for both unsqueezed and squeezed input light, and for cases where atomic collisions are considered or neglected. 
Firstly, Figs.~\ref{fig:colormap_phi_var}(a)-(d) shows the variation of $S^{\phi}(\omega)$ and $\zeta^{\phi}(\omega)$  with the detection angle $\phi$, for models with no collision, as compared to Figs.~\ref{fig:colormap_phi_var}(e)-(h), where collision or inter-atomic interactions are considered. From the numerical observations, it is evident that the dominant effect of collision is to shift the spectral peaks, which also causes the sensitivity to shift. This is true for both unsqueezed and squeezed light. Figures~\ref{fig:colormap_phi_var}(a) and (c), looks at the output spectrum and sensitivity for unsqueezed light, respectively. The maximum of the spectrum occurs at $\phi/\pi = 1/2$, which corresponds to the analytical obtained value. The sensitivity also is best close to the spectral peaks. However, as observed in Figs.~\ref{fig:colormap_phi_var}(b) and (d), the variation of $S^\phi$ is no longer straightforward and the maximal value occurs at a different detection angle $\phi$. The sensitivity again is optimal close to the spectral peaks.

The variation of the enhancement factor with respect to the homodyne detection angle $\phi$ and squeezing angle $\theta$, for different squeezing amplitudes $r$, are shown in Fig.~\ref{fig:zeta(phi_theta)}.
The enhancement factor increases with $r$, and by adjusting the detection angle the optimal sensitivity can be increased by a factor of 2.5. This is achieved for squeezing amplitude $r=2$, and the following squeezing and detection angles:  $(\theta,\phi)\in \{(2\pi/3,2\pi/3),(4\pi/3,\pi/3)\}$.

\section{BAE steady state beyond RWA\label{sec:A3}}

The typical approach to solve the equations of motion in the backaction evasion method is to find solutions in the good cavity regime, $\omega_m, G \gg\kappa$, as, for instance, discussed in Ref.~\cite{Woolley2013}. However, steady states of the Hamiltonian can also be obtained in the bad cavity regime, i.e., in the regime $\omega_m, G \ll \kappa$, and $\kappa\gg\Omega\gg\gamma$, under suitable circumstances. 
The Hamiltonian in Eq.~\eqref{bae-full-ham}, can be written in the interaction picture by transformation with $U=e^{-i\left(\omega_a a^{\dag}a + \omega_m (c^{\dag} c + d^{\dag} d)\right)t}$, such that
\begin{align}
    H'/\hbar&=\Omega (c^{\dag}c - d^{\dag}d) + \frac{G}{\sqrt{2}}\left(c e^{-i\omega_m t} + c^{\dag} e^{i\omega_m t}\right.\nonumber\\
    &\left. + d e^{-i\omega_m t} + d^{\dag} e^{i\omega_m t}\right)a^{\dag}a + \left(a\varepsilon^{*}_{+}e^{i(\delta-\omega_m) t}\right.\nonumber\\
    &\left. + a\varepsilon^{*}_{-}e^{-i(\delta-\omega_m) t} + h.c. \right).
\label{eq:H'}
\end{align}
Now, the Floquet expansion to the optical operator $a$, is given by 
$a = \sum_{j=-\infty}^{\infty} a_{j}e^{ij\delta t}$. Deriving the equation of motion for the expectation value of $a$, and using a mean-field or semiclassical approximation for correlations, the equations are given by
\begin{align}
\dot{a}&=\sum_{j=-\infty}^{\infty} (ija_{j} + \dot{a}_{j})e^{ij\delta t},\label{eq:substi}\\
&=-i\frac{G}{\sqrt{2}}\left(c e^{-i\omega_m t} + c^{\dag} e^{i\omega_m t} + d e^{-i\omega_m t} + d^{\dag} e^{i\omega_m t}\right)a\nonumber\label{eq:EOM_a}\\
&+ \varepsilon_{+} - \frac{\kappa}{2}a.
\end{align}
Comparing Eqs.~\eqref{eq:substi} and \eqref{eq:EOM_a}, and comparing coefficients of $e^{i\delta t}$ up to the first order and neglecting fluctuations, the relevant relations are
\begin{align}
&0~:~\dot{a}_{0}=-i\frac{G}{2}\left(ca_{1}+c^{\dag}a_{-1}+ da_{1} + d^{\dag}a_{-1}\right)-\frac{\kappa}{2},\\
&e^{i\delta t}:~i\delta a_{1}+\dot{a}_1 = -i\frac{G}{2}\left((c+d)a_2 + (c^{\dag} + d^{\dag})a_0\right) \nonumber\\
&~~~~~~~~~~~~~~~~~~~~ + ~\varepsilon_{+} -\frac{\kappa}{2}a_1,\\
&e^{-i\delta t}:-i\delta a_{-1}+\dot{a}_{-1} = -i\frac{G}{2}\left((c+d)a_0 + (c^{\dag} + d^{\dag})\right.\nonumber\\
&~~~~~~~~~~~~~~~~~~~~~~~~~~~\left. \times~ a_{-2}\right) + \varepsilon_{-} -\frac{\kappa}{2}a_{-1}.
\label{eq:general_EOMs_1}
\end{align}
The second-order terms are
\begin{align}
&e^{2i\delta t}:~2i\delta a_{2}+\dot{a}_{2}=-i\frac{G}{\sqrt{2}}(c^{\dag} + d^{\dag})a_{1} - \kappa a_{2},\\
&e^{-2i\delta t}:-2i\delta a_{-2}+\dot{a}_{-2}=-i\frac{G}{\sqrt{2}}(c + d)a_{-1} - \kappa a_{-2}.
\label{eq:general_EOMs_2}
\end{align}
Assuming that the optical field is stationary and therefore setting all $\dot{a}_j=0$, the following expressions for the steady state are obtained, 
\begin{align}
    \bar{a}_{2}&=\frac{iG\langle c^{\dag} + d^{\dag}\rangle \bar{a}_1}{\kappa/2 - 2i\delta},~
    \bar{a}_{-2}=\frac{iG\langle c + d\rangle \bar{a}_{-1}}{\kappa/2 + 2i\delta},\\
    \bar{a}_{1}&=\frac{\varepsilon_{+} - iG\left(\langle c^{\dag} + d^{\dag}\rangle \bar{a}_0 + \langle c + d\rangle \bar{a}_2\right)}{\kappa/2 + i\delta},\\
    \bar{a}_{-1}&=\frac{\varepsilon_{-} - iG\left(\langle c^{\dag} + d^{\dag}\rangle \bar{a}_{-2} + \langle c + d\rangle \bar{a}_{0}\right)}{\kappa/2 - i\delta},\\
    \bar{a}_{0}&=-\frac{i\sqrt{2}G\left(\langle c + d\rangle \bar{a}_{1} + \langle c^{\dag} + d^{\dag}\rangle \bar{a}_{-1} \right)}{\kappa}.
\end{align}
In the bad cavity regime, $G,\omega_m\ll\kappa$, the steady state values of $\bar{a}_{\pm 2}$ is directly proportional to the factor $G/\kappa$, which can be significantly small. Therefore the terms $\bar{a}_{\pm 2}$ and higher orders can be neglected in this regime, and a much simpler solution can be achieved by assuming just the first order driving amplitudes and atomic sideband modes, such that
\begin{equation}
    \bar{X}_{c(d)}=\bar{X}^{(1)}_{c(d)}e^{i\delta t} + \bar{X}^{(-1)}_{c(d)}e^{-i\delta t} + \bar{X}^{(0)}_{c(d)}.
\end{equation}
Substituting the above first-order expanded operators $\bar{X}_{c(d)}$  in Eqs.~\eqref{eq:general_EOMs_1}-\eqref{eq:general_EOMs_2} and comparing the different orders, the relations are
\begin{align}
\bar{a}_{0}&=0,~~~\bar{X}^{(\pm 1)}_{c(d)} = 0,\\
\bar{X}^{(0)}_{c(d)}&=-\frac{G}{\left(\omega_m \pm \Omega\right)}\left(|a_0|^2 + |a_1|^2 + |a_{-1}|^2\right),\\
i\delta \bar{a}_{1}&=-\frac{\kappa}{2}\bar{a}_{1} + i\Omega_{\rm eff}\left(|a_{1}|^2 + |a_{-1}|^2\right)\bar{a}_{1} - i\varepsilon_{+},\label{a+}\\
i\delta \bar{a}_{-1}&=-\frac{\kappa}{2}\bar{a}_{-1} + i\Omega_{\rm eff}\left(|a_{1}|^2 + |a_{-1}|^2\right)\bar{a}_{-1} - i\varepsilon_{-},\label{a-}\\
\Omega_{\rm eff}&=\frac{G(\Omega^2_d + \Omega^2_c)}{\Omega^2_c \Omega^2_d + \mathcal{A}^2}~~{\rm or}~~\Omega_{\rm eff}=\frac{2G\omega_{m}}{\omega_{m}^2 - \Omega^2},
\end{align}
where the two values of $\Omega_{\rm eff}$ are for the cases with and and without inter-atomic interaction or collisions in the dynamics, respectively. 
Making the above substitutions, the steady-state populations $n_{\pm 1}=|\bar{a}_{\pm 1}|^2$ under the above approximations, can be solved using a set of coupled cubic equations
\begin{align}
n_1\left[\left(n_1 + n_{-1} - \frac{\delta}{\Omega_{\rm{eff}}}\right)^2 + \left(\frac{\kappa}{\Omega_{\rm{eff}}}\right)^2\right]&=\frac{\vert\varepsilon_{+}\vert^2}{\Omega_{\rm{eff}}^2},\nonumber\\
n_{-1}\left[\left(n_1 + n_{-1} + \frac{\delta}{\Omega_{\rm{eff}}}\right)^2 + \left(\frac{\kappa}{\Omega_{\rm{eff}}}\right)^2\right]&=\frac{\vert\varepsilon_{-}\vert^2}{\Omega_{\rm{eff}}^2}.
\label{eq:coup_cubic}
\end{align}
Note that the occupation numbers of the two steady-states can be different i.e., $n_{1}\neq n_{-1}$.

\begin{figure}[t]
\centering
\includegraphics[width=3.1in]{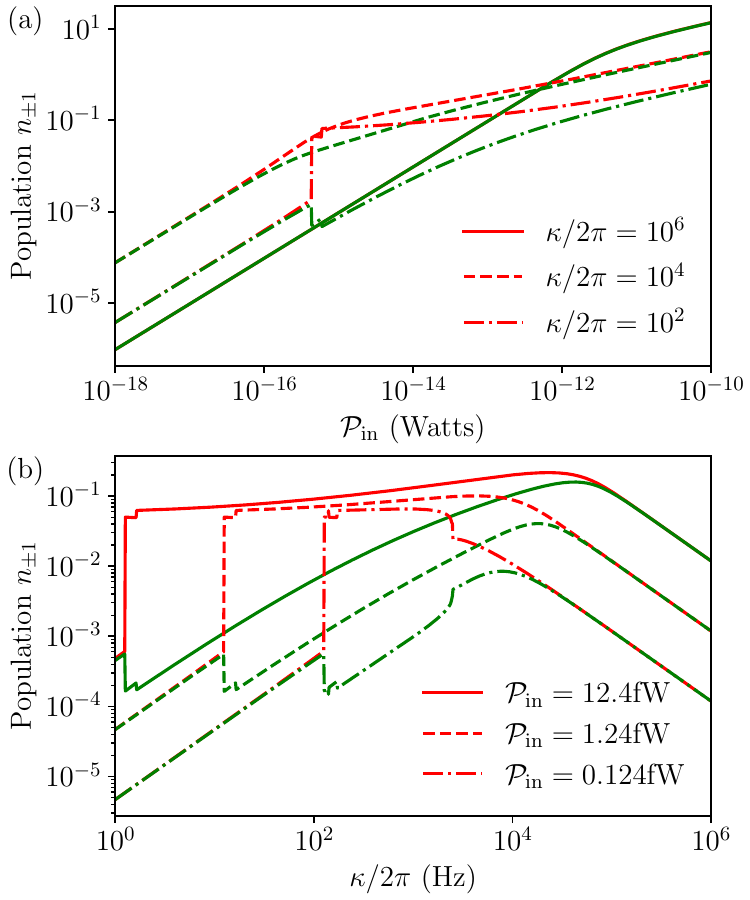}
\caption{Optical bistability in the steady state solutions of the equation of motion in BAE scheme, as a function of (a) input power $\mathcal{P}_{\rm in}$ and (b) cavity loss rate $\kappa$.
}
\label{fig:bistability}
\end{figure}

\begin{figure}[t]
\centering
\includegraphics[width=3.2in]{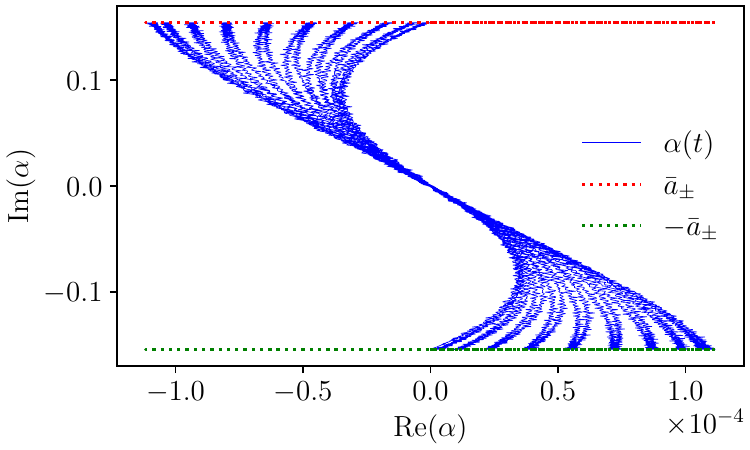}
\caption{Phase space evolution of the coherent state complex amplitude $\alpha(t)$, obtained numerically. The amplitude is constrained by the analytical bounds obtained by solving for the steady state equations.}
\label{fig:ss_bounds}
\end{figure}

The linearization of the equation of motion is based on the assumption that the operators can be written as the sum of a steady state and fluctuations. As such, it is necessary that a monostable steady state exists in the regime $G \ll \kappa$. 
Solving Eq.~\eqref{eq:coup_cubic} numerically exhibits the regions of bistability that appears as a sharp discontinuity in the sideband populations $n_{\pm}$ as the input power or the cavity dissipation rate is varied as shown in Fig.~\ref{fig:bistability}. For lower input power $\mathcal{P}_{\rm in}$, monostable region exists for most cavity dissipation. For higher input powers, to avoid bistability, it is sufficient to choose the cavity dissipation rate $\kappa/2\pi \geq 10^{5}$~Hz.

If the term $G$ is very small, the second term in the right hand side of Eqs.~\eqref{a+}-\eqref{a-} vanishes and the steady state solution reduces to 
$\bar{a}=a_s=\dfrac{|\varepsilon_{\pm}|}{\delta^2 + (\kappa/2)^2}$. 
This can then be considered to be the steady state value to linearize the optical operators in the BAE scheme.
The validity of the above analytical expression can be checked by numerically solving the time-dependent set of differential equations in the mean field regime obtained from Eq.~\eqref{eq:H'}. 
The time-dependent complex solution $\alpha$ of the optical field are bounded in the phase space under the steady state oscillation amplitude $|\bar{a}_{\pm}|$, as shown in Fig.~\ref{fig:ss_bounds}. This shows that the solutions of the differential equations in the phase-space is strictly bounded by the steady-state amplitudes, which can then be used 
to find the operable monostable regions where rotational sensitivity is to be measured.

\section{Spectral density coefficients for the BAE scheme\label{coeff_bae}}
The coefficients $\{A(\omega), B(\omega), C(\omega), D(\omega)\}$ in the expressions for the optical quadrature and the \com{spectral density} components in Eqs.~\eqref{eq:P_decompose}-\eqref{eq:spectrum_bae}, respectively, are obtained from the solutions in Eq.~\eqref{eq:full_solution1}-\eqref{eq:full_solution6}. 
These coefficients, with respect to the output optical quadrature $P{\rm out}(\omega)$, are
\begin{align}
    A^{0}(\omega)&=\sqrt{2}\kappa\chi^{2}_{a}(\omega)\bar{a}^2_{\pm}\left(\chi_c(\omega - \delta) + \chi_c(\omega+\delta) \right. 
    \nonumber\\
    &\left. +\chi_d(\omega - \delta) + \chi_d(\omega+\delta)\right),  
\end{align}
\begin{align}
    B^{0}(\omega)&=-\kappa\chi_a(\omega) -1,
    \\ \nonumber\\
    C^{\pm1}(\omega)&=\sqrt{2\kappa}G\bar{a}\chi_a(\omega \pm\delta)(\omega_m+\Omega)\chi_c(\omega),\\ \nonumber\\
    D^{\pm1}(\omega)&=\sqrt{2\kappa}G\bar{a}\chi_a(\omega \pm\delta)(\omega_m-\Omega)\chi_d(\omega),\\ \nonumber\\
    A^{\pm2}(\omega)&=\sqrt{2}\kappa G^2\bar{a}^2\chi_a(\omega \mp 2\delta)\chi_a(\omega)\left(\chi_c(\omega \mp\delta)\right.\nonumber\\ 
    &\left.+ \chi_d(\omega \mp \delta)\right),
\end{align}
where all higher-order coefficients are zero, which truncates the Floquet expansion at the second order.



\bibliography{citations}

\end{document}